\documentclass[manuscript]{aastex}

\usepackage{pdflscape}
\usepackage{epstopdf}
\usepackage{wasysym}
\usepackage{multirow}
\usepackage{verbatim}

\shorttitle{Synthetic spectral libraries for SDSS--III/APOGEE}
\shortauthors{Zamora et al.}

\begin{document}


\title{New H$-$band Stellar Spectral Libraries for the SDSS--III/APOGEE survey}


\author{O. Zamora\altaffilmark{1,2},  D. A.
Garc{\'{\i}}a-Hern{\'a}ndez\altaffilmark{1,2}, C. Allende Prieto\altaffilmark{1,2}, R.
Carrera\altaffilmark{1,2},  L. Koesterke\altaffilmark{3}, B. Edvardsson\altaffilmark{4},
F. Castelli\altaffilmark{5}, B. Plez\altaffilmark{6}, D. Bizyaev\altaffilmark{7,8}, K. Cunha\altaffilmark{9}, A. E.
Garc{\'{\i}}a P\'erez\altaffilmark{1,2}, B. Gustafsson\altaffilmark{4}, J. A.
Holtzman\altaffilmark{8}, J. E. Lawler\altaffilmark{10}, S. R. Majewski\altaffilmark{11}, A.
Manchado\altaffilmark{1,2,12}, Sz. M\'esz\'aros\altaffilmark{13}, N. Shane\altaffilmark{14}, M. Shetrone\altaffilmark{15}, V. V.
Smith\altaffilmark{16}, G. Zasowski\altaffilmark{17}}

%F. Castelli\altaffilmark{4}
\altaffiltext{1}{Instituto de Astrof{\'{\i}}sica de Canarias, E-38205 La Laguna, Tenerife, Spain}
\altaffiltext{2}{Departamento de Astrof{\'{\i}}sica, Universidad de La Laguna (ULL), E-38206 La Laguna, Tenerife, Spain}
\altaffiltext{3}{Texas Advanced Computing Center, The University of Texas at Austin, Austin, TX 78759, USA}
\altaffiltext{4}{Department of Physics and Astronomy, Division of Astronomy and Space Physics, Box 515, SE-751 20 Uppsala, Sweden}
\altaffiltext{5}{Istituto Nazionale di Astrofisica, Osservatorio Astronomico di Trieste, via Tiepolo 11, I-34143 Trieste, Italy}
\altaffiltext{6}{Laboratoire Univers et Particules de Montpellier, Universit\'e de Montpellier, CNRS, F$-$34095 Montpellier, France}
\altaffiltext{7}{Apache Point Observatory, P.O. Box 59, Sunspot, NM 88349-0059, USA}
\altaffiltext{8}{New Mexico State University, Las Cruces, NM 88003, USA}
\altaffiltext{9}{Observat\'orio Nacional, S\~ao Crist\'ov\~ao, Rio de Janeiro, Brazil}
\altaffiltext{10}{Department of Physics, University of Wisconsin-Madison, 1150 University Avenue, Madison, Wisconsin 53706, USA}
\altaffiltext{11}{Department of Astronomy, University of Virginia, Charlottesville, VA 22904-4325, USA}
\altaffiltext{12}{Consejo Superior de Investigaciones Cientificas (CSIC)}
\altaffiltext{13}{ELTE Gothard Astrophysical Observatory, H-9704 Szombathely, Szent Imre herceg \'ut, Hungary}
\altaffiltext{14}{Department of Astronomy, University of Virginia, Charlottesville, VA 22904-4325, USA}
\altaffiltext{15}{University of Texas at Austin, McDonald Observatory, Fort Davis, TX 79734, USA}
\altaffiltext{16}{National Optical Astronomy Observatories, Tucson, AZ 85719, USA}
\altaffiltext{17}{Department of Physics and Astronomy, Johns Hopkins University, Baltimore, MD 21218, USA}

\email{ozamora@iac.es}

%\and

%\author{R. J. Hanisch\altaffilmark{5}}
%\affil{Space Telescope Science Institute, Baltimore, MD 21218}


%\altaffiltext{1}{Visiting Astronomer, Cerro Tololo Inter-American Observatory.
%CTIO is operated by AURA, Inc.\ under contract to the National Science
%\altaffiltext{2}{Society of Fellows, Harvard University.}


\begin{abstract}
The Sloan Digital Sky Survey--III (SDSS--III) Apache Point Observatory Galactic
Evolution Experiment (APOGEE) has obtained high resolution (R $\sim$ 22,500), high
signal-to-noise ratio ($>$ 100) spectra in the H$-$band ($\sim$1.5$-$1.7
$\mu$m) for about 146,000 stars in the Milky Way galaxy. We have computed spectral
libraries with effective temperature ($T\rm{_{eff}}$) ranging from 3500 to 8000 K
for the automated chemical analy\-sis of the survey data. The libraries, used to
derive stellar parameters and abundances from the APOGEE spectra in the SDSS--III
data release 12 (DR12), are based on ATLAS9 model atmospheres and the
ASS$\epsilon$T spectral synthesis code. We present a second set of libraries based
on MARCS model atmospheres and the spectral synthesis code Turbospectrum. The
ATLAS9/ASS$\epsilon$T ($T\rm{_{eff}}$ = 3500$-$8000 K) and MARCS/Turbospectrum
($T\rm{_{eff}}$ = 3500$-$5500 K) grids cover a wide range of metallicity ($-$2.5
$\leq$ [M/H] $\leq$ $+$0.5 dex), surface gravity (0 $\leq$ log $g$ $\leq$ 5 dex),
microturbulence (0.5 $\leq$ $\xi$  $\leq$ 8 km~s$^{-1}$), carbon ($-$1 $\leq$
[C/M] $\leq$ $+$1 dex), nitrogen ($-$1 $\leq$ [N/M] $\leq$ $+$1 dex), and
$\alpha$-element ($-$1 $\leq$ [$\alpha$/M] $\leq$ $+$1 dex) variations, having
thus seven dimensions. We compare the ATLAS9/ASS$\epsilon$T and
MARCS/Turbospectrum libraries and apply both of them to the analysis of the
observed H$-$band spectra of the Sun and the K2 giant Arcturus, as well as to a
selected sample of well-known giant stars observed at very high-resolution. The
new APOGEE libraries are publicly available and can be employed for chemical
studies in the H$-$band using other high-resolution spectrographs. 
\end{abstract}


\keywords{astrochemistry -- radiative transfer -- stars: atmospheres -- surveys}



\section{Introduction}

Previous near-infrared (JHK bands) spectroscopic observations of individual or
small selected samples of giant stars have been limited in scope and mostly
biased towards the brightest sources. This has prevented the study of the
chemical abundance patterns in unbiased samples of stars towards the inner
(and dusty) parts (e.g., Galactic bulge and center) of our Galaxy, significantly
hampering progress towards a full understanding of the formation and chemical
(and dynamical) evolution of the Milky Way. This unfortunate situation has
dramatically changed in the new era of massive high-resolution spectroscopic
surveys. In particular, the Apache Point Observatory Galactic Evolution
Experiment (APOGEE) has focused on collecting high-resolution and high-quality
H$-$band spectra for a large ($>$ 10$^{5}$ stars) sample of giant stars, with
access to the inner - and more extinguished - regions of our Galaxy.

APOGEE is  one of the four spectroscopic surveys of the Sloan Digital Sky Survey
III (SDSS--III; e.g., Eisenstein et al. 2011). It is a high-resolution (R
$\equiv$ $\lambda$/$\Delta\lambda$ = 22,500) H$-$band (1.514--1.696 $\mu$m)
spectroscopic survey spanning all stellar populations in our Galaxy (see
e.g., Allende Prieto et al. 2008; Eisenstein et al. 2011; Majewski et al. 2015).
During the period from 2011 to 2014, the APOGEE survey collected about 500,000
spectra of $\sim$146,000 stars, predominantly post-main sequence stars (red
giants, subgiants, and red clump stars), using the Sloan Foundation 2.5m telescope
(Gunn et al. 2006) and an innovative multi-object IR spectrograph (Wilson et
al. 2010). High-resolution stellar spectra of red giants in the H$-$band show a
rich diversity of absorption lines from a wide variety of atoms and molecules,
with OH, CN, and CO the most important molecular contributors. To ascertain the
stellar atmospheric parameters and measure chemical abundances from the observed
spectra, the APOGEE Atmospheric Stellar Parameters and Chemical Abundances
Pipeline (ASPCAP; Garc{\'{\i}}a P\'erez et al. 2015) relies on an algorithm that
identifies the best-fitting synthetic spectrum for each observed spectrum. The
fitting code uses interpolation over pre-computed multi-dimensional grids of
synthetic spectra (i.e., model stellar spectral libraries) to find the best model
(with the minimum $\chi^2$ values) for each observed spectrum. Synthetic spectra
are calculated using classical model atmospheres (see e.g., M\'esz\'aros et al.
2012) and extensive atomic and molecular line lists (Shetrone et al. 2015). 

The SDSS--III APOGEE public data release 10 (DR10; Ahn et al. 2014) was based on
synthetic spectral libraries computed using Castelli \& Kurucz (2003) model
atmospheres (see, e.g., M\'esz\'aros et al. 2013). The Castelli \& Kurucz
(2003) model atmospheres incorporate line opacity by means of opacity distribution
functions (ODF) and are based on solar (or scaled solar) chemical abundances from
Grevesse \& Sauval (1998). For the APOGEE main survey targets, the fundamental
stellar parameters (effective temperature ($T\rm{_{eff}}$), surface gravity (log
$g$), and metallicity ([M/H]))\footnote{A linear relationship between
microturbulence and surface gravity is adopted (see Garc{\'{\i}}a P\'erez et al.
2015); e.g., $\xi$ = 2.478 $-$ 0.325 $\times$ log~$g$ is used in DR12 (Alam et al.
2015).} and the relative abundances of $\alpha$-elements ([$\alpha$/M]; in this
case O, Mg, Si, S, Ca, and Ti), carbon ([C/M]), and nitrogen ([N/M]) were released
in DR10. The final SDSS--III APOGEE public data release, DR12 (Alam et al. 2015)
is  based on self-consistent spectral libraries, where the same chemical
abundances are used both in the computation of the model atmospheres and in the
spectral synthesis. Moreover, newer solar reference abundances from Asplund et al.
(2005) are now adopted \footnote{The helium reference abundance adopted for
the computation of the synthetic spectral libraries is: 12 + $\log_{10}$(N$_{\rm
He}$/N$_{\rm H}$) = 10.93, where N$_{\rm He}$ and N$_{\rm H}$ are the number
density of helium and hydrogen nuclei, respectively.}.  In addition to the main
stellar parameters the individual element abundances for up to 15 elements
(typically with a precision of 0.1 dex  or better) are also released in DR12. 

In this paper we present for the first time the H$-$band stellar spectral
libraries for the SDSS--III APOGEE survey\footnote{The stellar spectral libraries
are available online;
\url{data.sdss3.org/sas/dr12/apogee/spectro/redux/speclib/asset/kurucz\_filled/solarisotopes/}},  which can be used
as well for chemical studies using other high-resolution spectrographs working in
the H$-$band. The DR12 spectral libraries are based on ATLAS9 model atmospheres 
(see M\'esz\'aros et al. 2012) and calculated with the ASS$\epsilon$T (Koesterke
et al. 2008; Koesterke 2009) spectral synthesis code. In addition to the official
family of ATLAS9/ASS$\epsilon$T DR12 spectral libraries, we have computed similar
spectral libraries based on MARCS model atmospheres, with the Turbospectrum
synthesis code (Alvarez \& Plez 1998; Plez 2012). We provide a comparison between
these different model atmospheres and spectral synthesis codes in order to check
the validity of the adopted DR12 synthetic spectral libraries. In Section
\ref{libraries}, we describe the parameter range and the calculation method of the
ATLAS9/ASS$\epsilon$T and MARCS/Turbospectrum stellar spectral libraries, while
Section 3 discusses systematic differences between the two grids of synthetic
spectra. Both grids are applied to the observed H$-$band spectra of the Sun and
the K2 giant Arcturus in Section 4. In Section 5, we use both model stellar
spectral libraries to derive the chemical patterns in a selected sample of
well-known giant stars observed with the Fourier Transform Spectrograph (FTS) 
on the Kitt Peak National Observatory 4m Mayall reflector. Our main conclusions
and future work are given in Section 6.

\section{Synthetic Spectral Libraries for APOGEE} \label{libraries}

As mentioned above, the SDSS--III APOGEE DR12 results (Alam et al. 2015) are based
on  ATLAS9/ASS$\epsilon$T synthetic spectral libraries. The APOGEE synthetic
spectral libraries are continuously improved and will be updated in the future
(see Section 6). In this section, we describe two H$-$band stellar spectral
libraries developed for the SDSS--III APOGEE survey, based on ATLAS9 and MARCS
model atmospheres and computed with the ASS$\epsilon$T and Turbospectrum spectral
synthesis codes, respectively. All MARCS/Turbospectrum computations were performed
on the Condor cluster at the Instituto de Astrof{\'{\i}}sica de Canarias (IAC).
Condor (or HTCondor)  is a High Throughput Computing (HTC) system developed by the
University of Wisconsin-Madison
(UW-Madison)\footnote{\url{http://research.cs.wisc.edu/htcondor/}}. In the case of
the IAC, Condor consists of a computer cluster with 808 CPUs. On the other hand,
the ATLAS9/ASS$\epsilon$T calculations were performed on the clusters Stampede and
Maverick operated by the Texas Advanced Computing Center (TACC).

It is to be noted here that the ATLAS9 and MARCS model atmospheres do not include
the line opacity for the polyatomic carbon molecules C$_{2}$H$_{2}$ and C$_{3}$
(M\'esz\'aros et al. 2012). These molecules are known to dominate the
infrared spectra of cool ($T\rm{_{eff}}$ $<$ 4000 K) carbon stars with C/O $>$
1.0, strongly affecting their thermal atmospheric structure. Thus, at present the
ATLAS9/ASS$\epsilon$T and MARCS/Turbospectrum synthetic spectra with
$T\rm{_{eff}}$ $<$ 3500$-$4000 K and C/O $>$ 1.0 are not reliable (see Section 6).

\subsection{ATLAS9/ASS$\epsilon$T spectral library} \label{atlas9}

The APOGEE ATLAS9/ASS$\epsilon$T spectral library makes use of the new ATLAS9 grid of
model atmospheres presented in M\'esz\'aros et al. (2012). ATLAS9 model atmospheres
(Kurucz 1993) are one-dimensional plane-parallel models computed in LTE and using opacity
distribution functions (ODFs) to handle line opacity (see e.g., Kurucz 2005). The
mixing-length scheme for convective energy transport is adopted, and the Kurucz atomic and
molecular line lists\footnote{\url{http://kurucz.harvard.edu/linelists.html}} are used
(see M\'esz\'aros et al. 2012 for more details). 

The new ATLAS9 model
atmospheres\footnote{\url{http://www.iac.es/proyecto/ATLAS-APOGEE/}} are based
on the recent solar composition by Asplund et al. (2005). This has the advantage
of matching the solar composition adopted also in the construction of the MARCS
models described in the next Section. M\'esz\'aros et al. (2012) carried out new
ODFs and Rosseland mean opacity calculations for several microturbulent
velocities ($\xi$ = 0, 1, 2, 4, and 8 km~s$^{-1}$), but all the APOGEE ATLAS9
model atmospheres are computed with $\xi$ = 2 km~s$^{-1}$. 

For the purpose of building the libraries, we divided the ATLAS9 model
atmospheres into two grids (GK and F) depending on the effective temperature
($T\rm{_{eff}}$), with the GK- and F-classes covering the 3500$-$6500 K and
5500$-$8000 K $T\rm{_{eff}}$ ranges, respectively. It is important to note that
missing (non-converged) ATLAS9 model atmospheres  are substituted by the nearest
model in chemical space at the same $T\rm{_{eff}}$ and log $g$.   The
non-converged ATLAS9 model atmospheres were a total of 6217 structures,
most of them corresponding to cool high surface gravity models 
($T\rm{_{eff}}$ $<$ 4000 K and log~$g$ $>$ 4.0) and cool C-rich models
($T\rm{_{eff}}$ $<$ 4000 K and C/O $>$ 0.75 ).

Most stars targeted by SDSS--III APOGEE are red giant and dwarf candidates, which
makes the ATLAS9 GK-class model grid the most important for APOGEE. The ATLAS9
GK-class models cover eleven $T\rm{_{eff}}$ and eleven log $g$ values (from
3500 to 6000 K in steps of 250 K and from 0 to 5 dex in steps of 0.5 dex,
respectively), seven metallicities [M/H]\footnote{The overall metallicity
[M/H] accounts for all elements with atom number Z $>$ 2 and   [M/H] =
$\log_{10}(N_{\rm M} / N_{\rm H})_{\star} -  \log_{10}(N_{\rm M} / N_{\rm
H})_{\odot}$, where N$_{\rm M}$ and N$_{\rm H}$ are the number density of any Z
$>$ 2 element and hydrogen nuclei, respectively. For the construction of
the synthetic spectral libraries,  all metals other than C, N, and
$\alpha-$elements are scaled with the corresponding [M/H] value.} (from $-$2.5 to
$+$0.5 dex in steps of 0.5 dex), and 81 values for the abundance of C and the
$\alpha$-elements ([C/M] and [$\alpha$/M] from $-$1.0 to $+$1.0 dex in steps of
0.25 dex). The $\alpha$-elements in the ATLAS9 models are as in the
introduction. Thus, 68,607 GK-class ATLAS9 models are present in the GK
ATLAS9 APOGEE grid used to produce the ATLAS9/ASS$\epsilon$T spectral library. 

The ATLAS9 F-class grid is used for the analysis of a much smaller number of
warmer stars such as the telluric line standards observed by the SDSS--III APOGEE
survey. The F-class ATLAS9-APOGEE grid contains also 68,607 model
atmospheres that cover $T\rm{_{eff}}$ values from 5500 K to 8000 K (in steps of
250 K) and the same [M/H], [C/M], and [$\alpha$/M] ranges as the GK-class one.

The code ASS$\epsilon$T was used to compute synthetic spectra for the Kurucz
ATLAS9 models. ASS$\epsilon$T was originally written for calculating spectra for
3D hydrodynamical simulations, and later a 1D branch was developed for dealing
with large numbers of classical hydrostatic model atmospheres. The code has an
option to pre-calculate the opacity as a function of temperature and a second
thermodynamical quantity (density or electron density) on a grid covering the
range of the models of interest, interpolating on-the-fly for solving the
radiative transfer equation. Opacity interpolations were avoided in the
calculations described above in order to maximize accuracy.
ASS$\epsilon$T adopts
the same software package used by Synspec/Tlusty (Hubeny \& Lanz 1995; Hubeny
2006) for computing opacities, and the input data described by Allende Prieto
(2008). Continuum opacities are mainly from the Opacity Project (Cunto et al.
1993) and the Iron Project (Nahar 1995; Bautista 1997), and line opacities are
shared in the calculations described in this paper for Turbospectrum/MARCS.

The opacities used in ASS$\epsilon$T are largely independent from those in Kurucz'
model atmospheres, especially in the case of continuum (bound-free and free-free) opacity.
We have, nevertheless, checked that the absolute fluxes predicted in the optical and
near-infrared are fairly similar for ASS$\epsilon$T and ATLAS9 for a solar-like star, where
H$^-$ dominates the continuum opacity (as for most APOGEE targets), while in the
near-UV the differences become larger due to different photoionization cross-sections for
metals (mainly iron, but also magnesium and other atoms). The overall equation
of state is also similar for the two codes, since when chemical equilibrium calculations
are performed in ASS$\epsilon$T, the final electron density for solar-like stars is
consistent with the original from the ATLAS9 model at a level of a few percent. In all
the ATLAS9 DR12 grids the electron density was iterated in ASS$\epsilon$T
to be consistent with  the atomic and molecular species considered in the synthesis.
The calculations for a few models with extreme compositions did result in iterated
electron densities that shifted the $\tau_{\lambda} \sim 100$ layer outside of the
ATLAS9 structure. ASS$\epsilon$T is set to stop when that occurs, so those calculations
were successfully completed relaxing that limit to $\tau_{\lambda} \sim 10$ .
 
The ATLAS9 APOGEE GK- and F-class model atmosphere grids mentioned above are used
in conjunction with the DR12 atomic/molecular line lists (linelist 201312161124;
Shetrone et al. 2015) to compute ATLAS9/ASS$\epsilon$T synthetic spectra with five
microturbulence values ($\xi$ = 0.5, 1, 2, 4, and 8 km~s$^{-1}$), five N
abundances ([N/M] from $-$1.0 to $+$1.0 dex in steps of 0.5 dex), and the solar C
isotopic ratio ($^{12}$C/$^{13}$C=89). Thus, in the end the GK- and F-class
ATLAS9/ASS$\epsilon$T grids contain seven dimensions (7D; $T\rm{_{eff}}$, log $g$,
$\xi$, [M/H], [C/M], [N/M], and [$\alpha$/M]) and the same number
(1,715,175) of total synthetic spectra. The model parameters for the
ATLAS9/ASS$\epsilon$T stellar spectral libraries are summarized in Table 1.
Finally, the synthetic spectra are smoothed to the APOGEE resolution,
continuum-normalized, re-sampled, and transformed to vacuum wavelengths. The
ATLAS9/ASS$\epsilon$T spectral libraries are then compressed using Principal
Component Analysis method (PCA, Pearson 1901) to prepare it for use by ASPCAP (see
Garc{\'{\i}}a P\'erez et al. 2015 for further details). We note that the line
spread function (LSF) for the ATLAS9/ASS$\epsilon$T DR12 spectral library is a
combo of those derived from several APOGEE fibers; actually an average of five
fibers LSFs characterized with a Gauss-Hermite function of variable resolution
with wavelength (see also Garc{\'{\i}}a P\'erez et al. 2015 for further details).
However, to make a fair comparison with the other spectral libraries treated in
this paper and  to avoid also systematic effects, we are using an
ATLAS9/ASS$\epsilon$T spectral library version that has been smoothed with a
Gaussian kernel to the APOGEE spectral resolution ($R$ = 22,500),  as the
MARCS/Turbospectrum one.

\subsection{MARCS/Turbospectrum spectral library} \label{marcs}

The APOGEE MARCS/Turbospectrum spectral library is based on the most recent
MARCS model atmospheres (Gustafsson et al. 2008). Basically, MARCS model
atmospheres are one-dimensional models in hydrostatic equilibrium that are
calculated in LTE and adopting the mixing-length theory for convection (Henyey
et al. 1965). The MARCS models are spherical for surface gravities log $g$
$\leq$ 3, while they are plane-parallel at higher gravities
(appropriate for dwarf stars; see Gustafsson et al. 2008). Line opacities are
treated with opacity sampling (OS).

The grid of MARCS model atmospheres for APOGEE was presented in M\'esz\'aros et
al. (2012)\footnote{The MARCS model atmospheres can be found in the MARCS Web
site; http://marcs.astro.uu.se/.} and we refer the reader to this paper and
Gustafsson et al. (2008) for more details. MARCS models covering nine
$T\rm{_{eff}}$ values (from 3500 to 5500 K in steps of 250 K)\footnote{We note
that the MARCS models with $T\rm{_{eff}}$ = 3700 K are used instead of those with
3750 K, because the $T\rm{_{eff}}$ step in the MARCS grid is 100 K below 4000 K.},
eleven log $g$ values (from 0 to 5 dex in steps of 0.5 dex)\footnote{Note that
plane-parallel and spherical MARCS model atmospheres are computed with a
microturbulence ($\xi$) of 1 and 2 km~s$^{-1}$, respectively (see, e.g.,
M\'esz\'aros et al. 2012).}, seven metallicities [M/H] (from $-$2.5 to $+$0.5 dex
in steps of 0.5 dex), and 25 combinations of C and $\alpha$-element
abundances\footnote{The $\alpha$-elements in MARCS are O, Ne, Mg, Si, S, Ar, Ca, and
Ti.} ([C/M] and [$\alpha$/M]
from $-$1.0 to $+$1.0 dex in steps of 0.5 dex) are present in the MARCS-APOGEE
grid used to construct the corresponding MARCS/Turbospectrum spectral library.
This results in a grid of $\sim$17,325 MARCS models; 1,062 models, however, do not
converge and are therefore missing in the MARCS-APOGEE grid. These non-converged
MARCS model atmosphere structures are replaced with the nearest model in chemical
space at the same $T\rm{_{eff}}$ and log $g$ (see below).  

Synthetic spectra in the APOGEE spectral range were generated with the
Turbospectrum package (Alvarez \& Plez 1998; Plez 2012), which shares the same
input data and radiative transfer routines with MARCS (Gustafsson et al.
2008). Turbospectrum is a 1D LTE spectral synthesis code that accounts for 600
molecules and uses the treatment of collisional line broadening described by
Anstee \& O'Mara (1995) and Barklem et al. (2000). It allows the computation of
flux (or intensity) synthetic spectra for stars of spectral type F and cooler. We
used version 13.1 of Turbospectrum but modified by us to use van der Waals
broadening (the Barklem treatment was used when possible, otherwise we used  van
der Waals constants from Kurucz; see Shetrone et al. 2015 for more details) for
the atomic lines\footnote{Version 14.1 of Turbospectrum includes these changes and
it is publicly available at \url{http://ascl.net/1205.004}.}. The synthetic
spectra are computed for an array of standard air wavelengths with a wavelength
step of 0.03 \AA. By using the MARCS-APOGEE grid (see above) and the DR12
atomic/molecular line lists (Shetrone et al. 2015) as input in Turbospectrum, we
constructed MARCS/Turbospectrum synthetic spectra with microturbulent velocities
of $\xi$ = 0.5, 1, 2, 4, and 8 km~s$^{-1}$ and with varying N content ([N/M] from
$-$1.0 to $+$1.0 in steps of 0.5 dex) and solar $^{12}$C/$^{13}$C ratio. This
resulted in a MARCS/Turbospectrum grid with seven dimensions (7D; $T\rm{_{eff}}$,
log $g$, $\xi$, [M/H], [C/M], [N/M], and [$\alpha$/M]) containing $\sim$ 43,375
synthetic spectra; the model parameters for the MARCS/Turbospectrum stellar
spectral library are also given in Table 1. It is to be noted here that in some
cases ($\sim$100) the spectral synthesis does not converge. However, these mainly
correspond to the most extreme and unrealistic abundance patterns (e.g., very high
or very low $\alpha$ element abundances). As in the case of the missing MARCS
model atmospheres mentioned above, we also replace the missing synthetic spectra
(fluxes) with the ones at the same $T\rm{_{eff}}$ and log $g$ and with the nearest
chemical composition. Finally, the MARCS/Turbospectrum synthetic spectra are
processed (i.e., smoothed, continuum-normalized, etc.) and PCA-compressed in the
same way as the ATLAS9/ASS$\epsilon$T library.

\section{Comparison between ATLAS9/ASS$\epsilon$T -- MARCS/Turbospectrum spectral libraries} \label{comp}

In this section we compare the Gaussian smoothed GK-ATLAS9/ASS$\epsilon$T and the
K-MARCS/Turbospectrum spectral libraries, which  since these grids overlap in
atmospheric parameters in the range 3500--5500 K.

Such comparisons can give indications on the uncertainties due to the use of
different model atmospheres, different spectral synthesis codes, as well as partly
different input data, with only the line lists being the same. It also allows to
check the possible influence of sphericity effects in the synthetic spectra (Sect.
\ref{sectcomp})\footnote{Sphericity effects are expected to be noticeable for low
gravity stars, as all ATLAS9 model atmospheres have plane-parallel geometry.} and
to explore possible systematic differences between both synthetic libraries in the
7D parameter space (Sect. \ref{sectsys}).

A total of 2,552 GK-ATLAS9/ASS$\epsilon$T test synthetic spectra 
were chosen to uniformly span the parameter space, 
while avoiding the grid boundaries; e.g., by
selecting one in ten consecutive spectra with different 7D parameters in the
nodes of the grid. Our synthetic spectra have surface gravities
from log~$g$ = 0.5 to 4.5 dex, effective temperatures $T\rm{_{eff}}$ from 3750
to 5250 K, microturbulent velocities from 1 to 4 km~s$^{-1}$, [C, N, $\alpha$/M]
from $-$0.5 to 0.5 dex, and [M/H] from $-$2.0 to 0.0 dex.

\subsection{Comparisons between spectral syntheses} \label{sectcomp}

For the study of possible differences (e.g.,sphericity effects) between both
families   of synthetic spectra, we extracted the same 2,552 test synthetic
spectra from the K-MARCS/Turbospectrum spectral library. The differences in the
synthetic spectra are estimated by deriving the root mean square difference
(r.m.s.) between the GK-ATLAS9/ASS$\epsilon$T and the K-MARCS/Turbospectrum pair
of spectral syntheses with the same 7D parameters. We performed comparisons for
two groups: low-gravity stars (log~$g$ $\leq$ 2.0) and high-gravity stars (log~$g$
$\geq$ 3.0). In this way, if sphericity effects are affecting the H$-$band
synthetic spectra of the low-gravity stars, we would find higher r.m.s. values for
this group.

\begin{landscape}
\begin{figure}
\epsscale{.80}
\begin{center}
%\plotone{thesun_16500-16600.eps}
\includegraphics[scale=0.7,angle=90]{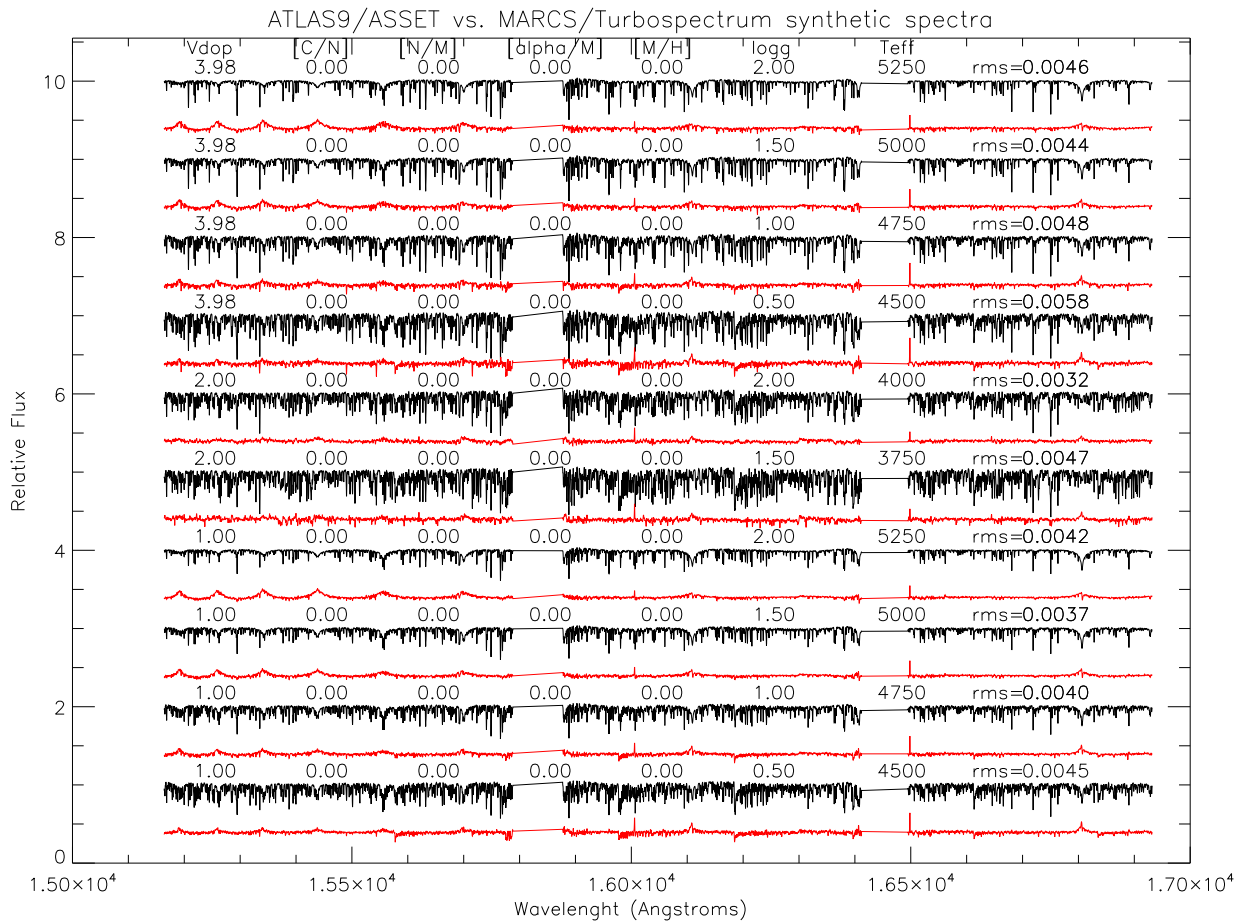}
\caption{Comparison between synthetic spectra extracted from the GK-ATLAS9/ASS$\epsilon$T and K-MARCS/Turbospectrum libraries
for low-gravity stars (log~$g \leq$ 2). The wavelength ranges covered by the three APOGEE detectors are showed.
Only ATLAS9/ASS$\epsilon$T spectra are displayed (in black). The residuals, computed
as MARCS $-$ ATLAS9 fluxes, have been multiplied by a factor five of to make them visible in the figure (red line).
The 7D parameters of each spectra are indicated above each spectrum (where $\xi$ $\equiv$ v$_{dop}$), together with the
root mean square (r.m.s.) value computed from each pair of
synthetic spectra with the same 7D parameters.
\label{atlasvsmarcslowg}}
\end{center}
\end{figure}
\end{landscape}

\begin{landscape}
\begin{figure}
\epsscale{.80}
\begin{center}
%\plotone{thesun_16500-16600.eps}
\includegraphics[scale=0.7,angle=90]{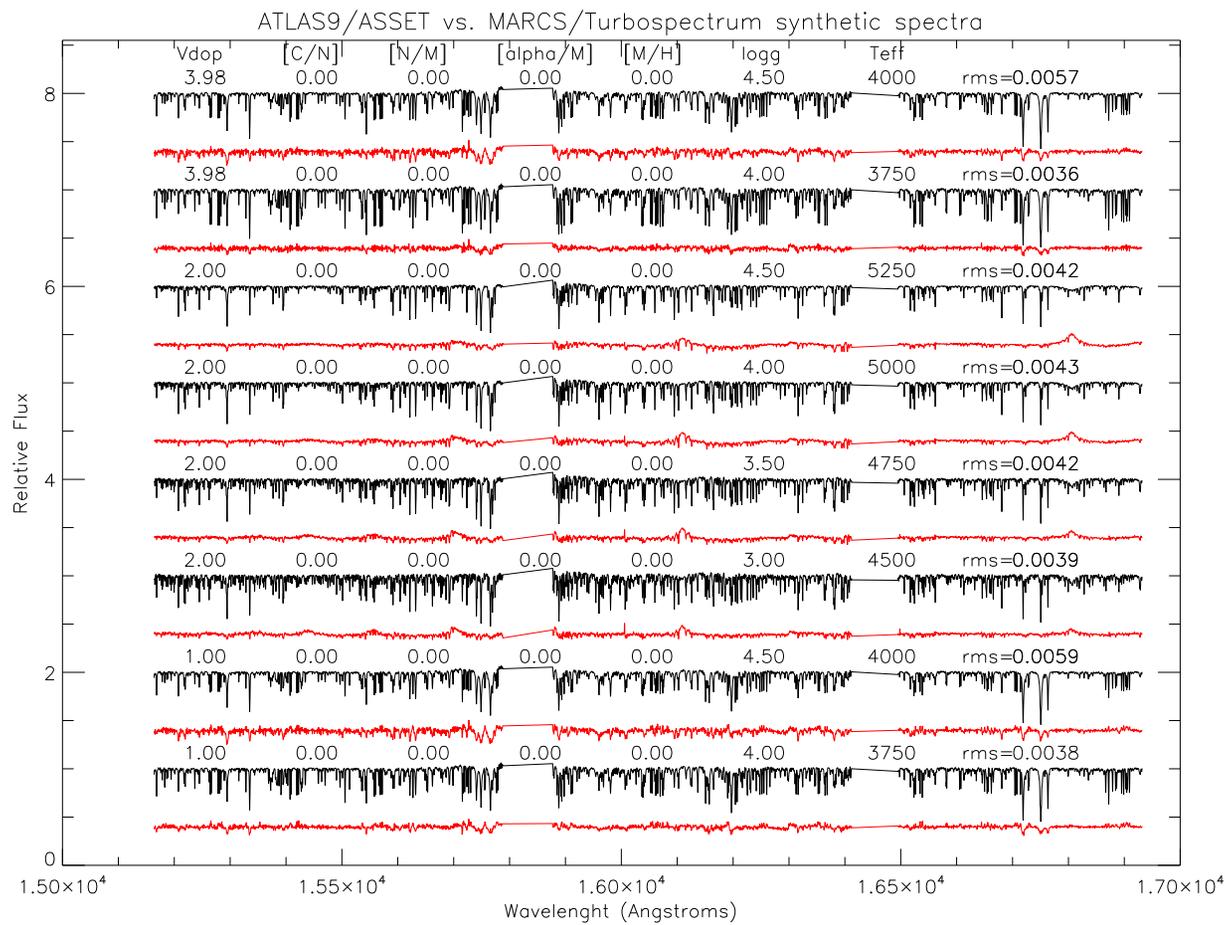}
\caption{Same as Figure 1 for log~$g$ $\geq$ 3.0.
\label{atlasvsmarcshighg}}
\end{center}
\end{figure}
\end{landscape}

Our results are summarized in Figs. \ref{atlasvsmarcslowg} and
\ref{atlasvsmarcshighg}, where we compare the r.m.s. values from some synthetic
spectra in our test sample; i.e., those with log~$g$ $\leq$ 2.0, [C/M] = 0.0,
[N/M]= 0.0, [$\alpha$/M] = 0.0 and [M/H] = 0.0 (Fig. \ref{atlasvsmarcslowg},
showing ten synthetic spectra in our test sample) and those with log~$g$ $\geq$
3.0, [C/M] = 0.0, [N/M]= 0.0, [$\alpha$/M] = 0.0 and [M/H] = 0.0 (Fig.
\ref{atlasvsmarcshighg}, displaying seven synthetic spectra in our test sample).
The r.m.s. values (of the order of $\sim$0.3-0.6 $\%$) are very similar for the
two groups. Interestingly, this indicates that the overall sphericity effects on
the chemical abundances are smaller in the H$-$band than in the optical wavelength
region (Heiter \& Eriksson 2006 found abundance differences as high as 0.35
dex); one reason may be that in the H$-$band we observe deeper atmospheric
layers. Indeed, we obtain similar r.m.s. values for other chemical compositions
([C/M], [N/M], [$\alpha$/M] and [M/H] values), validating the adoption of our
official DR12 library based on ATLAS9 models without any significant biases due to
the use of plane-parallel model atmospheres. The low r.m.s. values obtained are
indicative of no large differences between the computations based on
ASS$\epsilon$T and Turbospectrum spectral synthesis codes. We also note that the
(systematically)  deviating features in Figs. \ref{atlasvsmarcslowg} and
\ref{atlasvsmarcshighg} are hydrogen lines; this is because the spectral synthesis
codes (ASS$\epsilon$T and Turbospectrum) use different internal data for H. 
However, in the following section the H lines are not used in ASPCAP in 
fitting the best fit spectra.

\subsection{Systematic differences} \label{sectsys}

With the ultimate goal of exploring further the possible systematic differences
between both grids of APOGEE synthetic spectra, the 2,552 synthetic spectra in our
test sample, as extracted from the GK-ATLAS9/ASS$\epsilon$T spectral library, have
been fitted with the K-MARCS/Turbospectrum library using ASPCAP (see
Garc{\'{\i}}a P\'erez et al. 2015). In other words, we find out what
MARCS/Turbospectrum seven (7D) stellar parameters are recovered by the pipeline
when we treat the GK-ATLAS9/ASS$\epsilon$T synthetic spectra as input
\textit{observed spectra}. 

The results of this exercise are reported in Table \ref{tablesys} and in Figures 3
and 4. In Table \ref{tablesys} (columns 2 and 3) we show the median in the
difference $\Delta$ (= output MARCS parameter $-$ input ATLAS9 parameter) and the
dispersion ($\sigma$) of the differences obtained for the full sample.  To avoid
outliers from biasing the statistics, $\sigma$ is computed as the difference
between the maximum and minimum $\Delta$ after excluding the largest 15.85$\%$ of
the sample and the smallest 15.85$\%$, and divide it by two, which would
correspond to the standard deviation in a Normal distribution (see Fig. 3).  The
results of fitting the 2552 test synthetic spectra with the
GK-ATLAS9/ASS$\epsilon$T library are displayed in Table 2 (columns 4 and 5) and
Figure 4. In addition, we compare the ATLAS9 versus MARCS 7D parameters for two
different  sub-samples: low-gravity stars (log~$g$ $\leq$ 2.0.) and high-gravity
stars  (log~$g$ $\geq$ 3.0, see Table \ref{tablesys}; columns 6 to 9).

The MARCS/Turbospectrum spectral library (full sample) provides slightly higher
effective temperature, microturbulence, [N/M], and [$\alpha$/M], with median
values of 38.1 K, 0.02 dex, 0.09 dex, and 0.02 dex, respectively. On the other
hand, the MARCS/Turbospectrum library provides slightly lower metallicity and 
surface gravity than the ATLAS9/ASS$\epsilon$T one; median values of $-$0.03 dex
and $-$0.13 dex are found for [M/H] and log~$g$, respectively. Very similar carbon
abundances ([C/M]) are obtained with the MARCS/Turbospectrum spectral grid.  In
order to check if the differences in the 7D parameters mentioned above are
significant or merely the results of degeneracies, we compare the previous
results with the ones corresponding to the use of the ATLAS9/ASS$\epsilon$T
library as input (\textit{observed spectra}) but also running ASPCAP with the same
ATLAS9/ASS$\epsilon$T library. These results are reported in Table 2 (cols. 4 and
5), where we a find good consistency for log~$g$ and $T\rm{_{eff}}$, 
which have median values of 0.01 $\pm$ 0.04 dex and 3.10 $\pm$ 17.15 K,
respectively. We therefore conclude that the systematic differences between
MARCS/Turbospectrum and ATLAS9/ASS$\epsilon$T for log~$g$ and $T\rm{_{eff}}$ are
significant. The nitrogen abundance ([N/M]) may be slightly higher (median 0.09)
and is the most problematic parameter to recover for the MARCS/Turbospectrum
library, with the highest $\sigma$ of about 0.20 dex in [N/M] (see Table 2). This
N problem is not specific to the MARCS/Turbospectrum spectral library,
since the [N/M] parameter displays also the highest $\sigma$ if we just compare
the ATLAS9/ASS$\epsilon$T synthetic spectra with themselves (see Table 2 and
Fig. 4). We note that ASPCAP is limited in accuracy for low-metallicity
spectra ([M/H] $<$ $-$1.0), since the scarcity of lines in that regime causes
degeneracies among the stellar parameters (see Garc{\'{\i}}a P\'erez et al.
2015). Moreover, the $\sigma$ in N and in the other stellar parameters found
in tests by Garc{\'{\i}}a P\'erez et al. (2015) using ATLAS9/ASS$\epsilon$T
libraries are similar to the ones found here. 

If we compare the MARCS $-$ ATLAS9 residuals obtained for the 7D parameters  by
surface gravity groups, we find that high-gravity stars (log~$g$ $\geq$ 3.0)
display higher $\sigma$ values (for all parameters) than the low-gravity stars 
(log~$g$ $\leq$ 2.0, see Table 2; columns 6 to 9). However, the median parameter
values are quite similar with the exception of $T\rm{_{eff}}$, where the median
value for low-gravity stars is significantly higher (by about 60 K). The
differences in the 7D parameters between  the two log~$g$ groups become evident in
Figures 5 and 6. These figures show the residuals in surface gravity ($\Delta$
log~$g$) versus the residuals in the other stellar parameters  (from top to
bottom: [C/M], [N/M], [$\alpha$/M], [M/H], $\xi$, and $T\rm{_{eff}}$) for
low-gravity stars (Fig. 5) and high-gravity stars (Fig. 6). For low-gravity stars
(Fig. 5), the residuals for most of the data points are close to their median
values, with the exception of the already mentioned outliers (i.e., the largest
and smallest 15.85$\%$ of the sample) and the N problem. These outliers are
dominated by input ATLAS9/ASS$\epsilon$T  spectra of low-metallicity ([M/H] $<$
$-$1.0 dex). In spite of the outliers, the results for the MARCS/Turbospectrum and
ATLAS9/ASS$\epsilon$T spectral libraries in low-gravity stars are very similar,
and as mentioned above, the influence of possible sphericity effects on the
derived abundances using ATLAS9 plane-parallel model atmospheres is small in the
H-band. High-gravity outliers in this analysis correspond mainly to the
ATLAS9/ASS$\epsilon$T input synthetic spectra with [M/H] $<$ $-$1.0 dex. Further
work by the APOGEE ASPCAP team is needed to fully understand why ASPCAP
results degrate at high gravities.

\section{Comparisons of the Solar and Arcturus spectra with Syntheses from Spectral libraries} \label{sunarc}

In order to further investigate how consistent the SDSS--III APOGEE spectral 
libraries are, we have fitted the Sun and Arcturus observed spectra using the two
spectral libraries computed with both families of model atmospheres and spectral
syntheses codes  (ATLAS9/ASS$\epsilon$T and MARCS/Turbospectrum). We also compare
with the synthetic spectra obtained by using ATLAS9 model atmospheres and the
MOOG\footnote{\url{http://www.as.utexas.edu/$\sim$chris/moog.html}} spectral
synthesis code (Sneden 1973). This is relevant because ATLAS9/MOOG synthetic
spectra were used in the development of the DR12 line lists  (see Shetrone et al.
2015).  For the Sun, in particular, it is of interest to test how good our
spectral libraries perform comparing with a spectrum at much higher spectral
resolution than APOGEE's, and with well-known abundances. In addition, given
that APOGEE observes mostly giant stars, we can verify how well our
synthetic libraries reproduce the molecular lines (i.e., those suitable for CNO
abundance determinations) using the spectrum of a cooler giant star as Arcturus.

For the Sun, we use the high-resolution flux spectrum ($R$ = 400,000) by Livingston
$\&$ Wallace (1991). The Sun's synthetic spectra were computed adopting MARCS and
ATLAS9 model atmospheres with $T\rm{_{eff}}$ = 5777 K, log~$g$ = 4.4370, solar
composition by Asplund et al. (2005), and a microturbulent velocity of $\xi$ = 1.1
km~s$^{-1}$.  Three different synthetic spectra were computed with the following
model atmospheres and spectral synthesis codes: MARCS/Turbospectrum,
ATLAS9/ASS$\epsilon$T, and ATLAS9/MOOG.  The solar macroturbulent velocity was
taken into account in the synthetic spectra by convolving them with a Gaussian
profile having a FWHM of 1.58 km~s$^{-1}$ (Allende Prieto et al. 2001). The
synthetic spectra were also convolved with another Gaussian profile to match the
observed spectrum (FWHM = 1.87 km~s$^{-1}$)\footnote{The macroturbulence of
1.58 km~s$^{-1}$ is from optical spectra (Allende Prieto et al. 2001) and we need
an extra macroturbulence contribution to match the observed H-band solar
spectrum.}. Finally, the spectra were interpolated to the wavelengths (in air) of
the observed solar spectrum.  All three computed syntheses were compared with the
solar spectrum. The results from this comparison indicates a fairly good agreement
between the observed synthetic spectra, as well as a good agreement between the
synthetic spectra among themselves. The resulting global $\chi^2$
value for the Sun fitting were 15.05, 21.43 and 17.70 with the
MARCS/Turbospectrum, ATLAS9/ASS$\epsilon$T, and ATLAS9/MOOG synthetic spectra,
respectively\footnote{We estimate the error on the observed spectrum calculating
the standard deviation over a spectral region free of absorption lines and 
assuming this error constant along the full range in wavelength.}. Figure
\ref{sun} shows  the quality of the fits to the solar spectrum for the spectral
range 16500$-$16560 \AA, which includes several atomic and molecular lines. The
differences (or residuals) between the different types of synthetic spectra and
the Sun's observed spectrum are lower than 3$\%$ for most data points. The r.m.s.
value for the MARCS/Turbospectrum synthetic spectrum is slightly lower
($\sim$0.1$\%$) than  the ones for the ATLAS9/ASS$\epsilon$T and ATLAS9/MOOG
spectra. We note, however, that a perfect match between synthetic and observed
spectra  is not expected because of the convective line shifts and asymmetries in
the real stars, and of course because our modeling of the solar atmosphere is not
perfect.

In the case of the giant star Arcturus, we have used the FTS observed spectrum
smoothed to the resolution of APOGEE ($R$ = 22,500) and the best fit spectrum from
each library provided by ASPCAP/FERRE\footnote{FERRE is available from
\url{http://hebe.as.utexas.edu/ferre }.} (see also next section for further
details). The reduced global log~$\chi^2$ value for the fitting was
0.40 for the MARCS/Turbospectrum library and 0.41 for the ATLAS9/ASS$\epsilon$T
one (see Table 3 and Section 5).  We focus our comparisons of molecular features
in those spectral regions selected in the line-by-line abundance analysis by Smith
et al. (2013). These authors used four windows to extract the $^{12}$C abundance
from $^{12}$C$^{16}$O lines (15578$-$15586, 15774$-$15787, 15976$-$16000, and
16183$-$16196 \AA{}), four windows to extract the $^{16}$O abundance from
$^{16}$OH lines (15277$-$15282, 15390$-$15392, 15504$-$15507, and 16189$-$16193
\AA{}), and nine molecular lines of $^{12}$C$^{14}$N (15260, 15322, 15397, 15332,
15410, 15447, 15466, 15472, and 15482 \AA{}) to extract the abundance of 
$^{14}$N. 

The average of the residuals to fits to the observed Arcturus spectrum were
derived independently in each region/line for the two synthetic libraries. Very
small (r.m.s. $\sim$ 0.1 -- 0.3$\%$) differences are found between syntheses from
both libraries and the observed spectrum (and between the library synthetic
spectra themselves). Figures \ref{arcturus_C} and \ref{arcturus_NO} display the
fits in the $^{12}$C$^{16}$O and $^{16}$OH spectral regions and, in the
$^{12}$C$^{14}$N spectral lines mentioned above. In the $^{12}$C$^{16}$O windows,
the MARCS/Turbospectrum library spectrum fits slightly better the regions at
15578$-$15586 and 15976$-$16000 \AA{}, while the ATLAS9/ASS$\epsilon$T library
spectrum fits better the regions at 15774$-$15787 and 16183 $-$ 16196 \AA{}. For
the $^{16}$OH windows, the MARCS/Turbospectrum library spectrum fits better the
regions at 15277$-$15282 and 15390$-$15392 \AA{}, while the ATLAS9/ASS$\epsilon$T
library spectrum fits better the regions at 15504$-$15507 and 16189$-$16193 \AA{}.
Finally, for the $^{12}$C$^{14}$N lines, the MARCS/Turbospectrum library spectrum
fits slightly better the lines at 15260, 15397, 15466, and 15482 \AA{}, while the
ATLAS9/ASS$\epsilon$T library spectrum fits better the lines at 15322, 15332,
15447, and 15472 \AA{}. Both libraries provide just the same residual average for
the CN line at 15410 \AA{}. We conclude that both synthetic spectral libraries
provide an excellent fit to the molecular features in the spectrum of Arcturus;
the r.m.s. values between both synthetic libraries are no significant and of
the order of only $\sim$ 0.1 -- 0.3$\%$\footnote{Note that the f-values for atomic
lines were tuned to match the Sun and Arcturus but not those of molecular
transitions (see Shetrone et al. 2015).}.

\section{Application of the APOGEE spectral libraries to selected giant stars} \label{fts}

Given that most of the APOGEE sample are red giant stars, we are interested here
in exploring the abundance differences, as obtained by the APOGEE spectral
libraries, in a small sample of well known giant stars observed at very high
resolution. Smith et al. (2013) derived chemical abundances from a line-by-line
analysis of 15 elements in several well-known bright field giants and explored
what elements can be analyzed from APOGEE spectra. The sample analyzed here
consists in the four Smith et al. (2013) stars with $T\rm{_{eff}}$ $>$ 3500 K;
this includes two M-giants ($\beta$~And and $\delta$~Oph) and two K-giants
($\alpha$~Boo and $\mu$~Leo). For their study,  Smith et al. used high-resolution
spectra in the H--band acquired with the Fourier transform spectrometer
\citep[FTS; ][]{1979SPIE..172..121H} installed at the Coude focus of the Kitt Peak
National Observatory 4m Mayall reflector. The spectral resolution of these FTS
spectra varies from 45,000 to 100,000. The original spectra cover a wavelength
range larger than that of APOGEE, but \citet{2013ApJ...765...16S} restricted their
analysis to the spectral range from 1500 to 1700 nm. The infrared atlas spectrum
of $\alpha$~Boo (Arcturus) by \citet{1995PASP..107.1042H}, obtained with the same
instrument at a resolution of 100,000, is added to our sample of stars with
FTS spectra. The data were smoothed to the APOGEE resolution (i.e., $R$ =
22,500) by convolving with a Gaussian kernel. We also converted the resulting
convolved spectra to the APOGEE apStar FITS format described by Holtzman et al.
(2015). Our FTS sample includes the H$-$band spectra of $\alpha$~Boo and $\mu$~Leo
obtained with the APOGEE spectrograph, but using the New Mexico State University
1.0-meter Telescope (Holtzman et al. 2015; NMSU 1m, hereafter). The NMSU 1m
spectra are reduced with the APOGEE data reduction pipeline (Nidever et al. 2015).
All spectra mentioned above were processed with a quick version of ASPCAP,
\textit{QASPCAP}, which is short version of ASPCAP, and prepares the
spectra for the automated fitting with FERRE (see Garc{\'{\i}}a P\'erez et al.
2015 for a detailed description of ASPCAP). Finally, we derived the atmospheric
parameters and chemical abundances of 15 elements (see below) with FERRE,
interpolating in the ATLAS9/ASS$\epsilon$T and MARCS/Turbospectrum synthetic grids
described in this paper.

The atmospheric parameters and chemical abundances obtained are listed in
Tables~\ref{fts_parameters} and \ref{fts_abundances}, respectively. For all stars,
we find a very good agreement between the values obtained with both synthetic
grids. The log~$\chi^2$ of the fits for each spectral library are quite similar
(Table~\ref{fts_chi2}). The differences of the derived $T\rm{_{eff}}$, log $g$,
microturbulent velocity (in log scale), and [M/H] for each star (and synthetic
grid) are plotted in Figure~\ref{fig_fts_parameters}. The atmospheric parameters
derived here for the Arcturus giant are somewhat different from those adopted in
Smith et al. (2013): $\Delta$$T\rm{_{eff}}$ $<$ 90 K,  $\Delta$log $g$ $<$ 0.4
dex, $\Delta$$\xi$ $<$ 0.6 km~s$^{-1}$, $\Delta$[M/H] $<$ 0.2 dex.  In Smith et
al. (2013) the effective temperatures were based on the (J$-$K) color and derived
from an average of two calibrations: Gonz\'alez Hern\'andez \& Bonifacio (2009)
and Bessell et al. (1998). Their surface gravities were obtained from evolutionary
tracks and microturbulent velocities from the Fe I lines. The stellar parameters
derived here are purely spectroscopic. In addition, it is important to note that
Smith et al. (2013) carried out a line-by-line chemical abundance analysis using
the MOOG synthesis code and a so-called intermediate version of the APOGEE line
list (line list \textit{INT}; Shetrone et al. 2015) that is previous to the DR12
APOGEE line list used here.

Regarding abundances, we find in general very good agreement ($<$0.1 dex) between
the chemical abundances obtained by the ATLAS9/ASS$\epsilon$T and
MARCS/Turbospectrum stellar spectral libraries in the FTS stars (see
Fig.~\ref{fig_fts_abundances}). For Arcturus ($\alpha$ Boo), MARCS/Turbospectrum
performs slightly better  than ATLAS9/ASS$\epsilon$T, with the exception of N (see
below).  Despite the higher MARCS/Turbospectrum $T\rm{_{eff}}$ values, the set of
derived MARCS/Turbospectrum abundances in Arcturus (and also in $\mu$ Leo) are
slightly lower ($<$0.1 dex) than those from  ATLAS9/ASS$\epsilon$T because of the
generally lower metallicity and gravity obtained (which compensates the expected
abundance increase due to a higher $T\rm{_{eff}}$; Section 3) with the
MARCS/Turbospectrum library (see Tables~\ref{fts_parameters} and
\ref{fts_abundances}). However, N seems to be more affected and the
MARCS/Turbospectrum derived N abundances can be lower by 0.15 dex. On the other
hand, ATLAS9/ASS$\epsilon$T fits slightly better than MARCS/Turbospectrum in the
two FTS stars $\beta$ And and $\delta$ Oph; especially concerning nitrogen, where
MARCS/Turbospectrum gives N abundances lower by $\sim$0.2$-$0.3 dex (see
Fig.~\ref{fig_fts_abundances}). This, however, is likely due to the use of the
cooler 3700 K MARCS model in place of the 3750 K one, which is lacking in
the MARCS grid, to interpolate their exact matching value of effective
temperatures of about 3825--3850 K. 

In short, based on the results from the comparisons performed in this study, there
is good indication that the MARCS/Turbospectrum library, although with several
more models missing from the grid, gives results comparable to those from
the ATLAS9/ASS$\epsilon$T library \footnote{Garc{\'{\i}}a P\'erez et al. (2015)
used the same ATLAS9/ASS$\epsilon$T spectral library to analyze the spectra of FTS
stars but with a different order in the stellar parameters.}.

\section{Conclusions and future work} \label{conc}

We present the stellar spectral libraries for the final data release of the
SDSS--III APOGEE survey, which are used for the automated chemical analysis of
survey data. The spectral libraries employed in the data release 12 (DR12;
Alam et al. 2015) are constructed for a wide range in effective temperature
($T\rm{_{eff}}$ ranging from 3500 to 8000 K) and are based on ATLAS9 model
atmospheres and the ASS$\epsilon$T spectral synthesis code. We also present here a
second family of SDSS--III APOGEE stellar spectral libraries based on MARCS model
atmospheres and the Turbospectrum spectral synthesis code. The
ATLAS9/ASS$\epsilon$T ($T\rm{_{eff}}$ = 3500--8000 K) and MARCS/Turbospectrum
($T\rm{_{eff}}$ = 3500--5500 K) synthetic grids have seven dimensions (7D),
covering a wide metallicity ([M/H]), surface gravity (log $g$), microturbulence
($\xi$), carbon ([C/M]), nitrogen ([N/M]), and $\alpha$-element ([$\alpha$/M])
ranges of variation. 

We have compared both ATLAS9/ASS$\epsilon$T and MARCS/Turbospectrum spectral
libraries to a test sample of 2552 synthetic spectra with the same 7D stellar
parameters. The differences found between both families of synthetic spectra are
very small (r.m.s. values of the order of only $\sim$0.3$-$0.6 $\%$).
Interestingly, we find that the sphericity effects in the H$-$band seem to be
smaller than those previously found in the optical range and the ASS$\epsilon$T
and Turbospectrum spectral synthesis codes provide very similar  synthetic
spectra. By fitting the GK-ATLAS9/ASS$\epsilon$T library with the
K-MARCS/Turbospectrum library, we have found small systematic differences in the
seven main stellar parameters (7D; $T\rm{_{eff}}$, [M/H], log $g$, $\xi$, [C/M],
[N/M], and [$\alpha$/M]) automatically provided by the SDSS--III APOGEE survey for
low-gravity stars (log $g$ $\leq$ 2.0). The outliers correspond to low-metallicity
([M/H] $<$ $-$1.0) synthetic spectra. However, the results for high-gravity (log
$g$ $>$ 3.0) stars are worse than the former ones, and the average scatter
for the entire parameter space is higher than in low-gravity stars.
These outliers are also dominated  by synthetic spectra with [M/H] $<$ $-$1.0 dex.
Further work by the APOGEE ASPCAP team is needed to completely understand the
presence of these outliers in the 7D parameter space.

Both the DR12 SDSS--III APOGEE synthetic spectral library as well as the
additional spectral library based on the MARCS model atmospheres provide almost
identical model fits to the observed spectra of the Sun, Arcturus, and the stars
with FTS spectra. For example, they give an excellent fit to the Sun's spectrum as
well as to the molecular features (CO, OH, and CN) in the spectrum of Arcturus;
the differences (or residuals) between these synthetic libraries are of the order
of only $\sim$ 0.1 -- 0.3$\%$ (r.m.s.). We conclude that both SDSS--III APOGEE
synthetic spectral libraries provide very similar results (i.e., atmospheric
parameters and chemical abundances), which supports the use of the
ATLAS9/ASS$\epsilon$T synthetic grids (which otherwise cover a parameter space
much wider than the actual MARCS/Turbospectrum grid) in DR12. The SDSS--III APOGEE
synthetic spectral libraries presented here are publicly available online and they
can be used also for chemical analysis in the H$-$band making use of other
available high-resolution spectroscopic instruments working in the H$-$band.

The APOGEE stellar spectral libraries presented here will be improved for the
SDSS--IV/APOGEE--2 survey, and periodically updated in the future. We plan to
extend the MARCS/Turbospectrum stellar spectral library to cooler temperatures
(2500 $\leq$ $T\rm{_{eff}}$ $\leq$ 3500 K). The effect of other molecules such as
H$_{2}$O and FeH may be important at these extremely cool effective temperatures
and we will need to update the present APOGEE linelist by including these
molecules. Finally, we plan  to evaluate the effects of the missing
opacities for polyatomic molecules (like HCN, C$_{2}$H$_{2}$) on the structures
of the cool ($T\rm{_{eff}}$ $<$ 4000 K) and C-rich MARCS models atmospheres and we
plan to improve the latter C-rich models with new opacities for such polyatomic
molecules. 

\acknowledgments
The authors acknowledge the Texas Advanced Computing Center (TACC) at The University of
Texas at Austin for providing HPC resources that have contributed to the research results
reported within this paper (URL: http://www.tacc.utexas.edu). 
This paper made use of the IAC Supercomputing facility HTCondor 
(URL: http://research.cs.wisc.edu/htcondor).
O.Z., D.A.G.H., and A.M.
acknowledge support provided by the Spanish Ministry of Economy and Competitiveness under
grant AYA-2011$-$27754.
R.C. acknowledges support provided by the Spanish Ministry of Economy and Competiviness
under grants AYA2010$-$16717 and AYA2013$-$42781P.
Funding for SDSS--III has been provided by the Alfred P. Sloan
Foundation, the Participating Institutions, the National Science Foundation, and the U.S.
Department of Energy Office of Science. The SDSS--III web site is http://www.sdss3.org/.
SDSS--III is managed by the Astrophysical Research Consortium for the Participating
Institutions of the SDSS--III Collaboration including the University of Arizona, the
Brazilian Participation Group, Brookhaven National Laboratory, University of Cambridge,
Carnegie Mellon University, University of Florida, the French Participation Group, the
German Participation Group, Harvard University, the Instituto de Astrof{\'{\i}}sica de
Canarias, the Michigan State/Notre Dame/JINA Participation Group, Johns Hopkins
University, Lawrence Berkeley National Laboratory, Max Planck Institute for Astrophysics,
New Mexico State University, New York University, Ohio State University, Pennsylvania
State University, University of Portsmouth, Princeton University, the Spanish
Participation Group, University of Tokyo, University of Utah, Vanderbilt University,
University of Virginia, University of Washington, and Yale University.


%% in parentheses, after the keyword, but they will not be verified.

%{\it Facilities:} \facility{Nickel}, \facility{HST (STIS)}, \facility{CXO (ASIS)}.



\clearpage

\begin{landscape}
\begin{deluxetable}{llllllll}
\tabletypesize{\scriptsize}
\tablecaption{Parameters of the ATLAS9/ASS$\epsilon$T and MARCS/Turbospectrum spectral libraries}
\tablewidth{0pt}
\tabcolsep=0.1cm
\tablehead{
\colhead{Class} &
\colhead{$T_{\rm eff}$ range (step) }    & \colhead{log~$g$ range (step) } &
\colhead{$[$M$/$H$]$ range (step) } & \colhead{$[$C/M$]$ range (step) } & \colhead{$[$N/M$]$ range (step) } 
& \colhead{[$\alpha$/M] range (step) } & \colhead{log~$\xi$ range (step)\tablenotemark{a} }
\\
\colhead{} &
\colhead{(K)}    & \colhead{(dex)} &
\colhead{(dex)} & \colhead{(dex)} & \colhead{(dex)} &
 \colhead{(dex)} & \colhead{(km~s$^{-1}$)}
}
\startdata
ATLAS9/ASS$\epsilon$T &  & & & &  & & \\
\hline
\hline
GK & 3500$-$6000 (250) & 0$-$5 (0.5) & -2.5$-$0.5 (0.5) & -1.0$-$1.0 (0.25) & -1.0$-$1.0 (0.5) & -1.0$-$1.0 (0.25) & -0.301$-$0.903 (0.301) \\
F & 5500$-$8000 (250) & 0$-$5 (0.5) & -2.5$-$0.5 (0.5) & -1.0$-$1.0 (0.25) & -1.0$-$1.0 (0.5) & -1.0$-$1.0 (0.25) & -0.301$-$0.903 (0.301) \\
\hline
MARCS/Turbospectrum &  & &  & &  & & \\
\hline
\hline
K & 3500$-$5500 (250) & 0$-$5 (0.5) & -2.5$-$0.5 (0.5) & -1.0$-$1.0 (0.5) & -1.0$-$1.0 (0.5) & -1.0$-$1.0 (0.5) & -0.301$-$0.903 (0.301)\\
\enddata
\tablenotetext{a}{The microturbulence (log~$\xi$) step is given in log$_{10}$ units
(uniform step of 0.301 dex).  \label{parameters}}
\end{deluxetable}
\end{landscape}

\begin{deluxetable}{lcccccccc}
\tabletypesize{\tiny}
\tablecaption{Systematic differences between ATLAS9/ASS$\epsilon$T (input) and MARCS/Turbospectrum (output) spectral syntheses
in the 7D parameter space \label{tablesys}}
\tabcolsep=0.05cm
\tablewidth{0pt}
\tablehead{
\colhead{Parameter} & \colhead{Median(MARCS$-$ATLAS9)} &
\colhead{$\sigma$} & \colhead{Median(ATLAS9$-$ATLAS9)*} &
\colhead{$\sigma$} & \colhead{Median(MARCS$-$ATLAS9)} &
\colhead{$\sigma$} &
\colhead{Median(MARCS$-$ATLAS9)} &
\colhead{$\sigma$}
\\
\colhead{} & \colhead{Full sample} &
\colhead{} & \colhead{Full sample} &
\colhead{} & \colhead{log$~g$~$\leq$ 2.0 subsample} &
\colhead{} & \colhead{log$~g$~$\geq$ 3.0 subsample} &

}
\startdata
log~$\xi$       &0.02     &0.08 & 0.00 & 0.03  &0.01    &0.02 &0.08    & 0.19 \\
$[$C/M$]$       &0.00     &0.08 & 0.00 & 0.04  &0.00    &0.07 &$-$0.01 & 0.10 \\
$[$N/M$]$       &0.09     &0.20 & 0.01 & 0.15  &0.10    &0.17 &0.07    & 0.27 \\ 
$[\alpha/$M$]$  &0.02     &0.02 & 0.00 & 0.01  &0.03    &0.02 &0.01    & 0.03 \\
$[$M/H$]$       &$-$0.03  &0.04 & 0.00 & 0.01  &0.01    &0.04 &$-$0.04 & 0.04  \\
log~$g$         &$-$0.13  &0.15 & 0.01 & 0.04  &$-$0.11 &0.13 &$-$0.11  & 0.14  \\
$T\rm{_{eff}}$  &38.10    &50.15& 3.10 & 17.15  &64.30   &41.70 &20.00   & 59.20  \\
\enddata
\tablenotetext{*}{These are the results using synthetic spectra from the ATLAS9/ASS$\epsilon$T library as input and output for ASPCAP.}
\end{deluxetable}

\begin{deluxetable}{lcc}
\tabletypesize{\scriptsize}
\tablecaption{ATLAS9/ASS$\epsilon$T vs. MARCS/Turbospectrum log~$\chi^{2}$ values
in FTS stars\label{fts_chi2}}
\tablewidth{0pt}
\tablehead{
\colhead{Star} & \colhead{log~$\chi^{2}$ ATLAS9/ASS$\epsilon$T} &
\colhead{log~$\chi^{2}$ MARCS/Turbospectrum} 
}
\startdata
$\alpha$ Boo (FTS)   & 0.4075 &   0.3972 \\
$\alpha$ Boo (atlas) & 0.3524 &   0.3443 \\
$\alpha$ Boo (NMSU 1m)    & 2.0242 &   2.0236 \\ 
$\mu$ Leo  (FTS)     & 1.1852 &   1.1885 \\
$\mu$ Leo  (NMSU 1m)      & 1.8115 &   1.8260 \\
$\beta$ And  (FTS)\tablenotemark{a}   & 1.3842 &   1.3926 \\
$\delta$ Oph  (FTS)\tablenotemark{a}& 1.1643 &   1.1730 \\
\enddata
\tablenotetext{a}{These stars have atmospheric parameters corresponding to a hole in the MARCS/Turbospectrum grid.}
\end{deluxetable}

\begin{deluxetable}{lcccccccc}
\tabletypesize{\scriptsize}
\tablecaption{ATLAS9/ASS$\epsilon$T vs. MARCS/Turbospectrum stellar
parameters in FTS stars\label{fts_parameters}}
\tablewidth{0pt}
\tablehead{

\colhead{} & \colhead{} &\colhead{ATLAS9/ASS$\epsilon$T} &\colhead{} &\colhead{} &
\colhead{} & \colhead{MARCS/Turbospectrum} &\colhead{} &\colhead{} \\
\hline
\hline
\colhead{Star} &
\colhead{$T_{\rm eff}$} & \colhead{log~$g$} & \colhead{$[$M/H$]$} & \colhead{$\xi$} & 
\colhead{$T_{\rm eff}$} & \colhead{log~$g$} & \colhead{$[$M/H$]$} & \colhead{$\xi$}  
}
\startdata
$\alpha$ Boo (FTS)   & 4187 & 2.04 & -0.40 & 2.03 & 4192 & 1.85 & -0.47 & 2.07 \\
$\alpha$ Boo (atlas) & 4188 & 2.07 & -0.43 & 1.90 & 4192 & 1.95 & -0.48 & 1.92 \\
$\alpha$ Boo (NMSU 1m)    & 4208 & 2.07 & -0.50 & 1.15 & 4223 & 1.92 & -0.56 & 1.25 \\
$\mu$ Leo  (FTS)     & 4493 & 2.80 &  0.44 & 1.93 & 4520 & 2.76 &  0.40 & 1.99 \\
$\mu$ Leo  (NMSU 1m)      & 4560 & 2.98 &  0.36 & 1.00 & 4551 & 2.87 &  0.30 & 1.04 \\
$\beta$ And  (FTS)\tablenotemark{a}   & 3823 & 1.16 & -0.20 & 2.36 & 3791 & 1.19 & -0.24 & 2.42 \\
$\delta$ Oph  (FTS)\tablenotemark{a}  & 3832 & 1.45 &  0.03 & 2.21 & 3809 & 1.48 & -0.04 & 2.31 \\
 \enddata
 \tablenotetext{a}{These stars have atmospheric parameters corresponding to a hole in the MARCS/Turbospectrum grid.}
\end{deluxetable}

\begin{landscape}

\begin{deluxetable}{lcccccccccccccc}
\tabletypesize{\scriptsize}
\tablecaption{ATLAS9/ASS$\epsilon$T vs. MARCS/Turbospectrum element
abundances in FTS stars\label{fts_abundances}}
\tablewidth{0pt}
\tablehead{
\colhead{Star} &
\colhead{$[$Fe/H$]$} & \colhead{$[$C/H$]$} & \colhead{$[$N/H$]$} & \colhead{$[$O/H$]$} &
\colhead{$[$Mg/H$]$} & \colhead{$[$Al/H$]$} & \colhead{$[$Si/H$]$} &
\colhead{$[$K/H$]$} &
\colhead{$[$Ca/H$]$} &
\colhead{$[$Ti/H$]$} & \colhead{$[$V/H$]$} & \colhead{$[$Mn/H$]$} & \colhead{$[$Ni/H$]$} \\
}
\startdata
 & & & & & & & ATLAS9/ASS$\epsilon$T & & & & & & &\\
\hline
\hline
$\alpha$ Boo (FTS)   &-0.42 & -0.29 & -0.30 & -0.16 & -0.23 & -0.24 &  0.01 & -0.44 & -0.37 & -0.43 & -0.53 & -0.48 & -0.38\\
$\alpha$ Boo (atlas) &-0.45 & -0.31 & -0.35 & -0.20 & -0.25 & -0.37 & -0.10 & -0.45 & -0.42 & -0.43 & -0.58 & -0.49 & -0.39\\
$\alpha$ Boo (NMSU 1m)    &-0.52 & -0.46 & -0.58 & -0.39 & -0.39 & -0.27 & -0.30 & -0.51 & -0.51 & -0.26 & -0.71 & -0.55 & -0.32\\
$\mu$ Leo  (FTS)     & 0.44 &  0.43 &  0.91 &  0.49 &  0.36 &  0.50 &  0.53 &  0.50 &  0.23 &  0.62 &  0.30 &  0.50 &  0.50\\
$\mu$ Leo  (NMSU 1m)      & 0.31 &  0.37 &  0.73 &  0.37 &  0.22 &  0.41 &  0.33 &  0.19 &  0.21 &  0.64 &  0.32 &  0.50 &  0.47\\
$\beta$ And  (FTS)   &-0.21 & -0.35 &  0.20 & -0.11 & -0.07 & -0.12 & -0.06 & -0.30 & -0.26 & -0.07 & -0.23 & -0.11 & -0.19\\
$\delta$ Oph  (FTS)  & 0.00 & -0.08 &  0.30 &  0.08 &  0.06 &  0.19 &  0.24 &  0.27 & -0.08 &  0.17 & -0.04 &  0.21 &  0.03\\
\hline
 & & & & & & & MARCS/Turbospectrum & & & & & & &\\
\hline
\hline
$\alpha$ Boo (FTS)   &-0.49 & -0.37 & -0.38 & -0.17 & -0.26 & -0.24 & -0.06 & -0.45 & -0.37 & -0.45 & -0.56 & -0.53 & -0.43\\
$\alpha$ Boo (atlas) &-0.51 & -0.36 & -0.50 & -0.22 & -0.28 & -0.37 & -0.15 & -0.47 & -0.42 & -0.43 & -0.59 & -0.53 & -0.42\\
$\alpha$Boo (NMSU 1m)     &-0.59 & -0.50 & -0.64 & -0.39 & -0.40 & -0.26 & -0.36 & -0.52 & -0.51 & -0.35 & -0.72 & -0.58 & -0.37\\
$\mu$ Leo  (FTS)     & 0.38 &  0.40 &  0.91 &  0.47 &  0.32 &  0.50 &  0.41 &  0.50 &  0.23 &  0.58 &  0.30 &  0.50 &  0.50\\
$\mu$ Leo  (NMSU 1m)      & 0.24 &  0.31 &  0.73 &  0.32 &  0.20 &  0.41 &  0.27 &  0.15 &  0.18 &  0.57 &  0.26 &  0.50 &  0.41\\
$\beta$ And  (FTS)\tablenotemark{a}   &-0.27 & -0.44 & -0.08 & -0.34 & -0.03 & -0.29 & -0.02 & -0.36 & -0.38 & -0.28 & -0.36 & -0.15 & -0.23\\
$\delta$ Oph  (FTS)\tablenotemark{a}  &-0.07 & -0.16 &  0.07 & -0.11 & -0.11 & -0.02 &  0.27 &  0.19 & -0.19 & -0.08 & -0.15 &  0.17 & -0.02\\
\enddata
\tablenotetext{a}{These stars have atmospheric parameters corresponding to a hole in the MARCS/Turbospectrum grid.}
\end{deluxetable}

\end{landscape}

\begin{figure}
\epsscale{.80}
%\plotone{thesun_16500-16600.eps}
\begin{center}
\includegraphics[scale=0.52,angle=90]{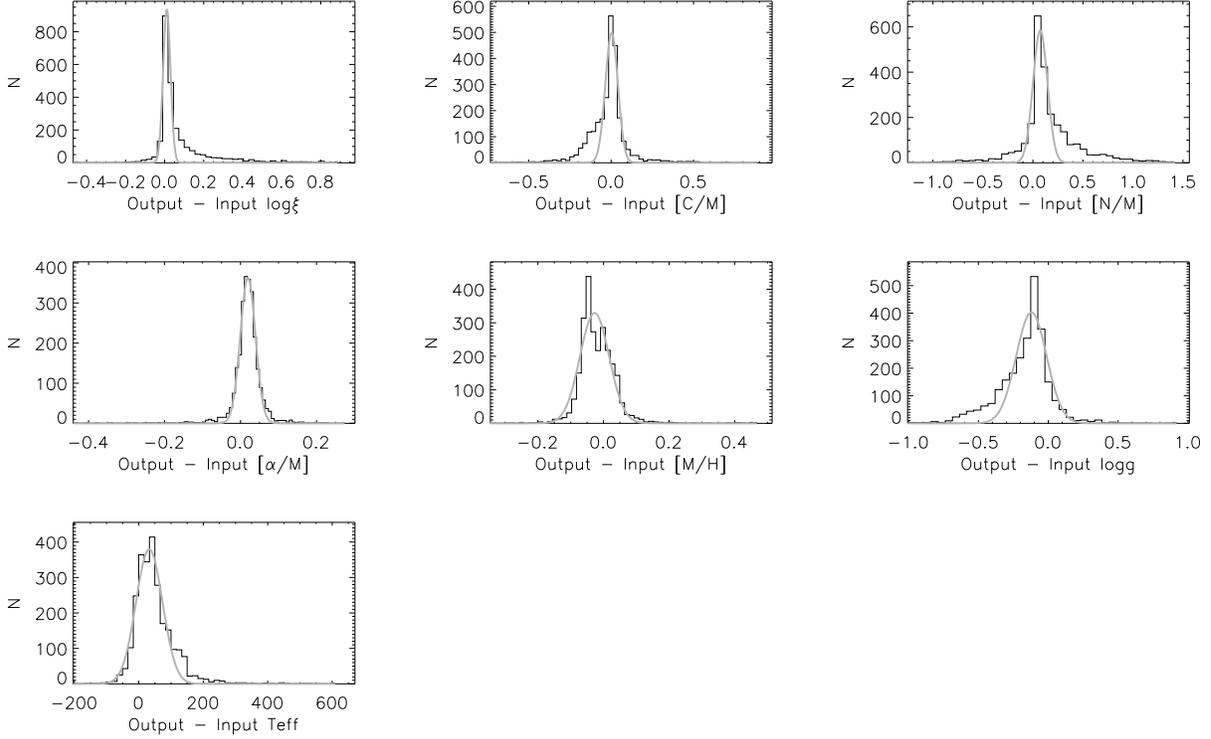}
\end{center}
\caption{Systematic differences between the ATLAS9/ASS$\epsilon$T and the MARCS/Turbospectrum
spectral libraries in the 7D parameter space.  The input stellar parameters are those from the
ATLAS9/ASS$\epsilon$T spectral library, while the output parameters are those derived/recovered
with the MARCS/Turbospectrum library by using ASPCAP. The distribution of the differences 
(i.e., output MARCS parameter $-$ input ATLAS9 parameter) is shown for the 7D grid (in black). A
Gaussian distribution fit, excluding the $\sim$15.85 $\%$ of the largest data differences  (i.e.,
outliers), is also displayed (grey curve). 
\label{sysparameters}}
\end{figure}

\begin{figure}
\epsscale{.80}
%\plotone{thesun_16500-16600.eps}
\begin{center}
\includegraphics[scale=0.52,angle=90]{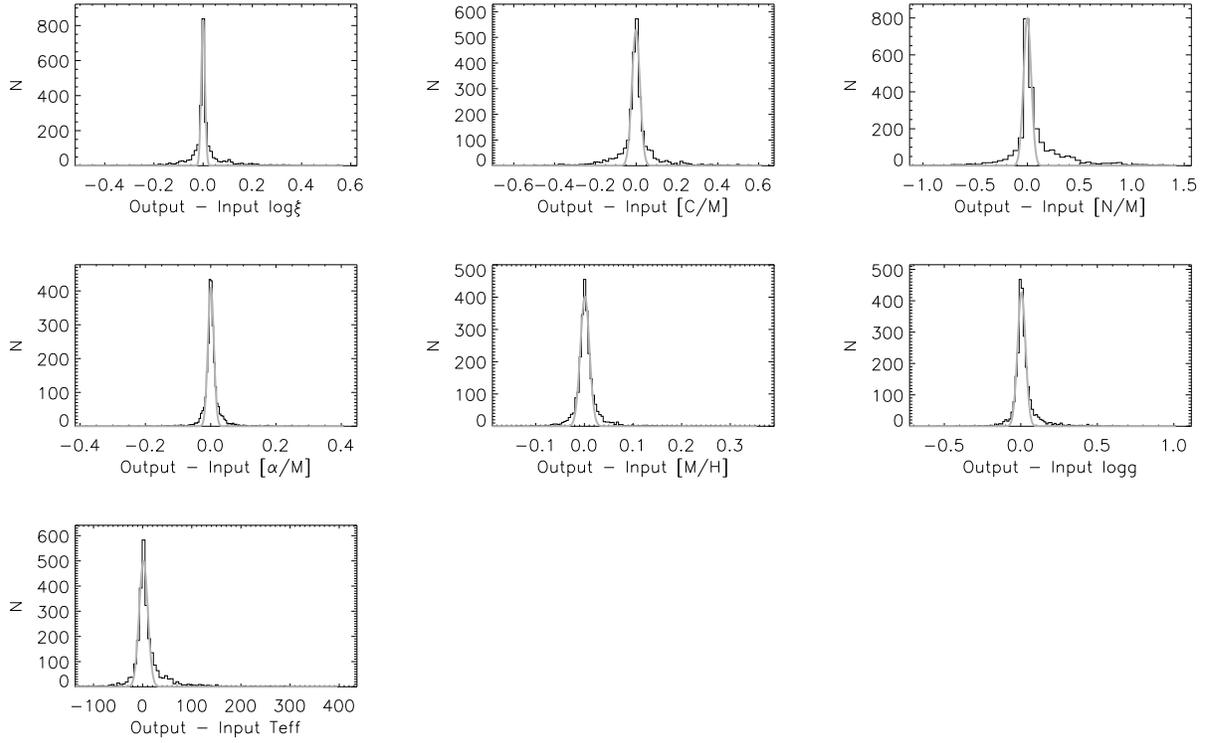}
\end{center}
\caption{Same as Figure 3 but comparing both the input and the output from the ATLAS9/ASS$\epsilon$T spectral library in the
7D parameter space. 
\label{sysparameters}}
\end{figure}

\begin{figure}
\epsscale{.80}
%\plotone{thesun_16500-16600.eps}
\begin{center}
\includegraphics[scale=0.7,angle=90]{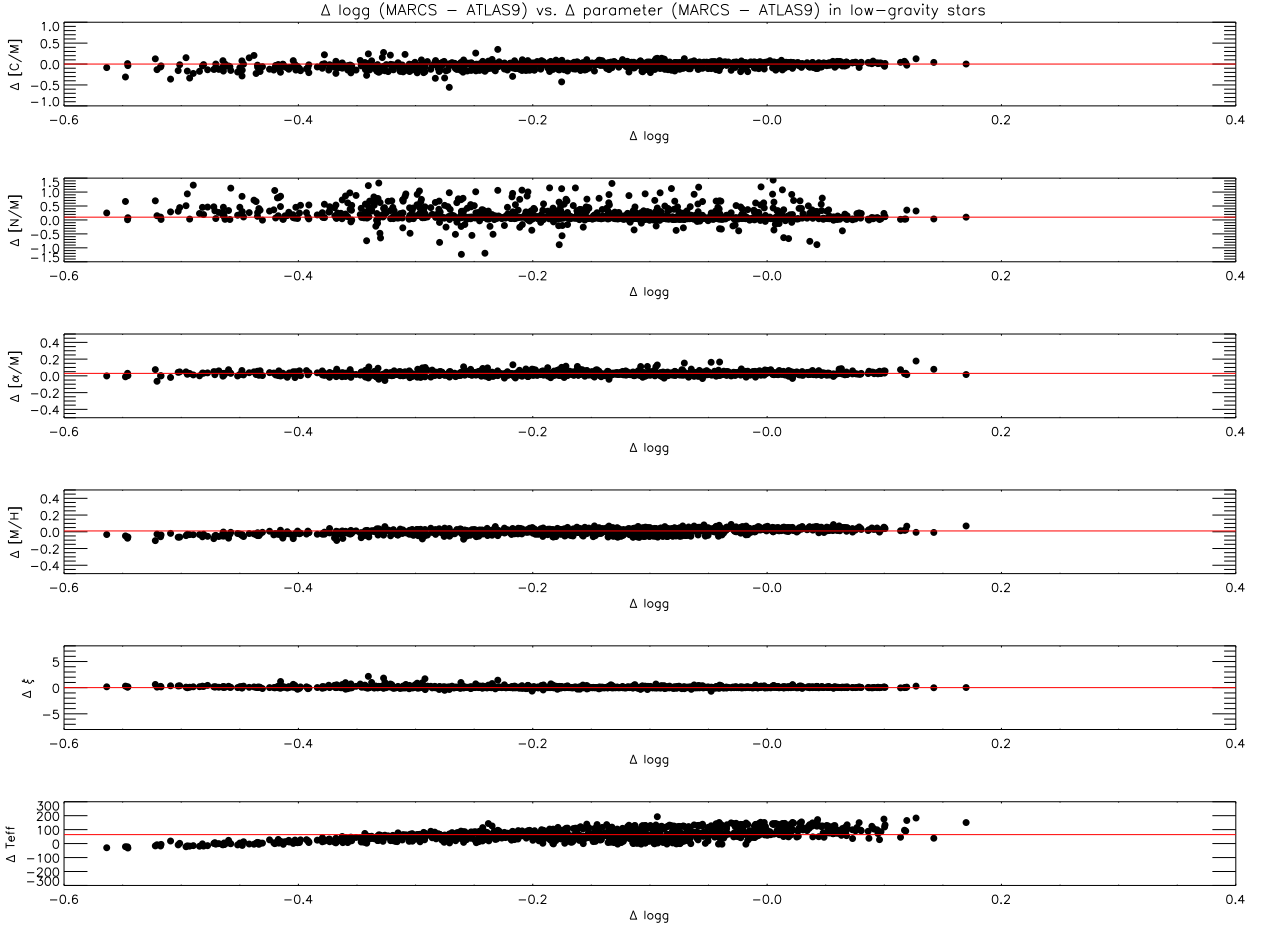}
\end{center}
\caption{Parameter differences obtained between MARCS/Turbospectrum and ATLAS9/ASS$\epsilon$T spectral libraries for
low-gravity (log~$g$ $\leq$ 2.0) stars. From top to bottom:
differences (MARCS $-$ ATLAS9) in carbon ([C/M]), nitrogen ([N/M]), $\alpha-$elements ([$\alpha$/M]), metallicity
([M/H]), microturbulent velocity ($\xi$), and effective temperature ($T\rm{_{eff}}$) versus MARCS $-$ ATLAS9 
differences in surface gravity ($\Delta$ log~$g$). The red
line indicates the median values of the parameter differences. \label{sysdifferencesdteff}}
\end{figure}

\begin{figure}
\epsscale{.80}
\begin{center}
%\plotone{thesun_16500-16600.eps}
\includegraphics[scale=0.7,angle=90]{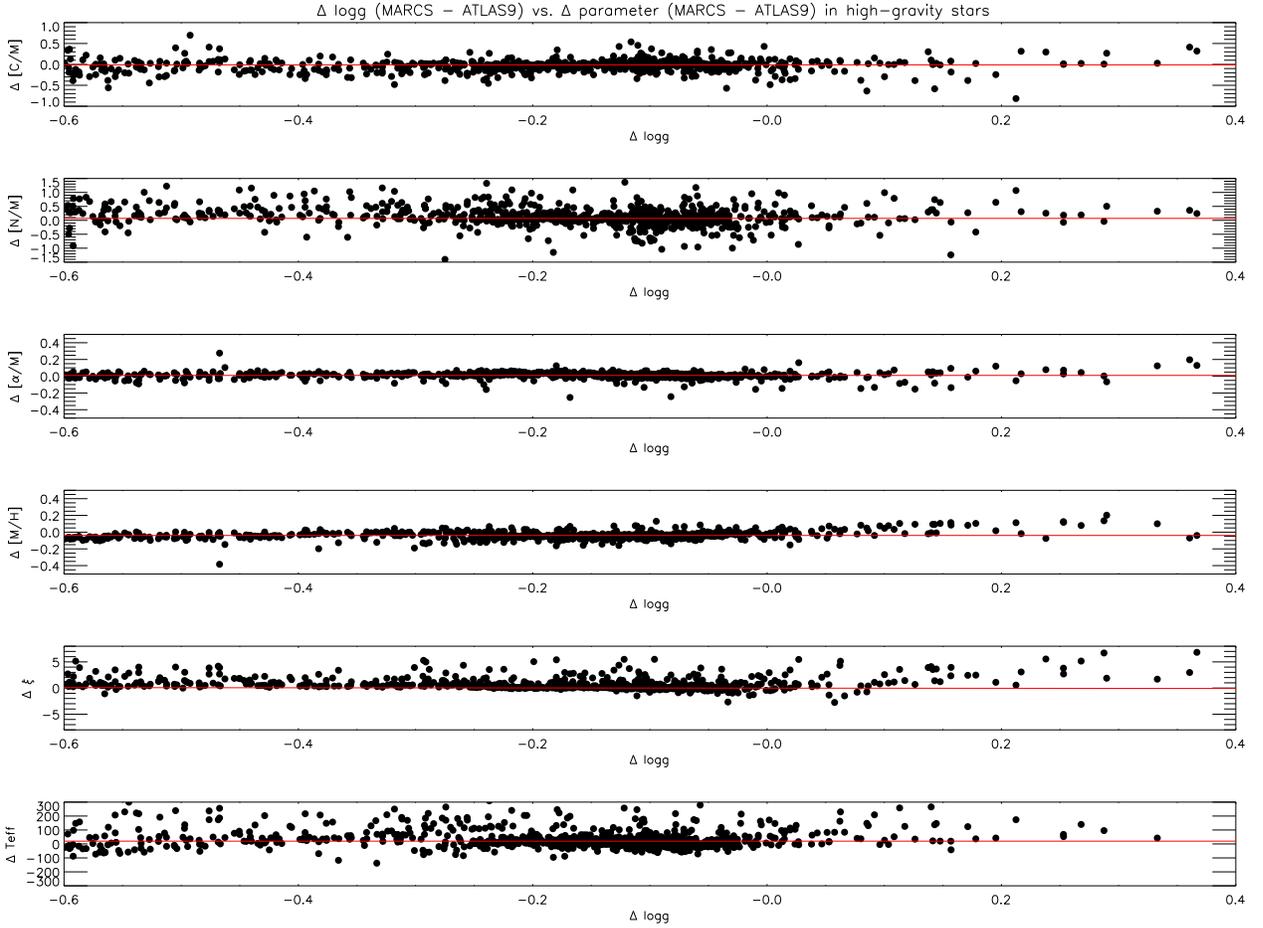}
\end{center}
\caption{Same as Figure 5 but for high-gravity stars; i.e., log~$g$ $\geq$ 3.0.
\label{sysdifferencesdg}}
\end{figure}

\begin{figure}
\epsscale{.80}
%\plotone{thesun_16500-16600.eps}
\includegraphics[scale=0.68,angle=90]{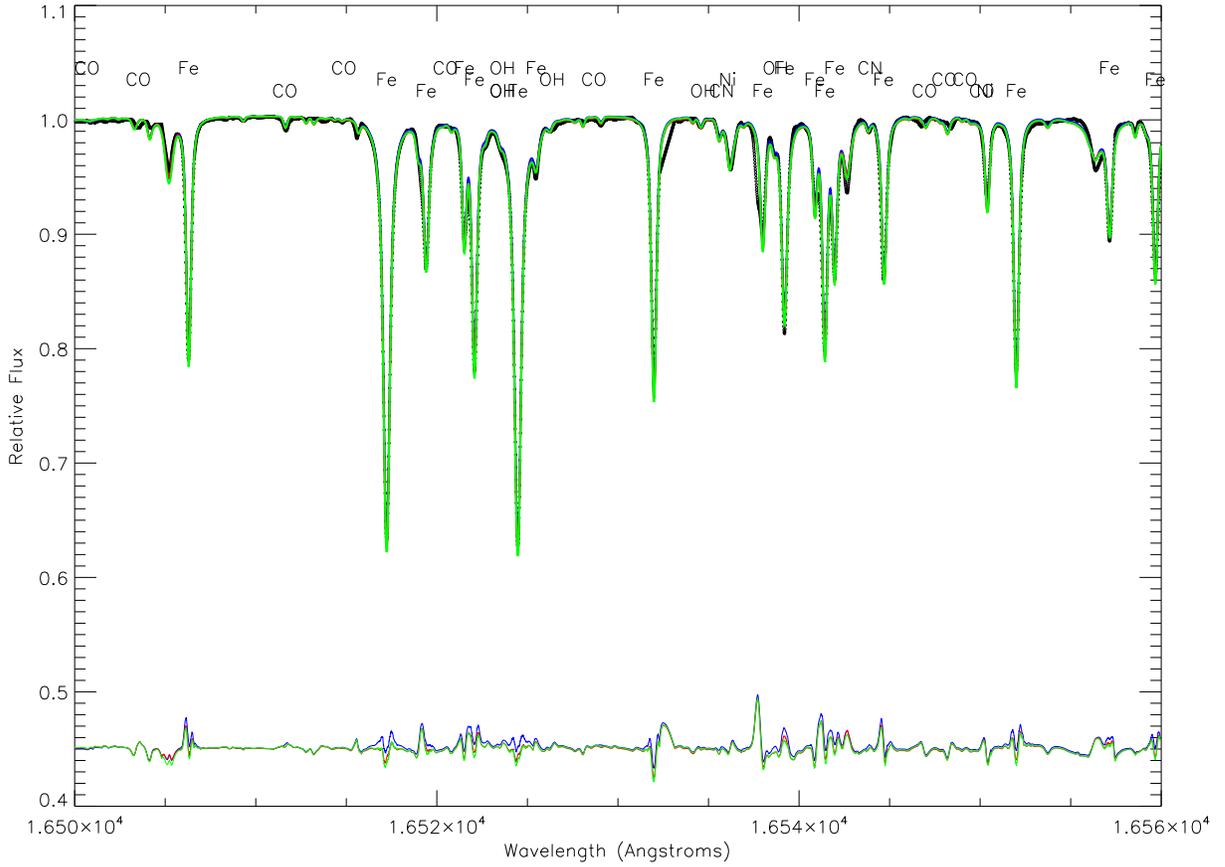}
\caption{High-resolution H$-$band observed spectrum of the Sun (black dots) in the
16500$-$16560 \AA~region versus the best fits obtained using
MARCS/Turbospectrum (red line), ATLAS9/ASS$\epsilon$T (blue line), and
ATLAS9/MOOG (green line) synthetic spectra. All spectra are expressed in air wavelengths. The residuals,
computed as flux(synthetic$-$observed)$+$0.45, are plotted at the bottom with
the same colors. The spectral features identified by Hinkle et al. (1995) are indicated
at the top.
\label{sun}}
\end{figure}
%\end{landscape}

\begin{figure}
\epsscale{.80}
%\plotone{thesun_16500-16600.eps}
\includegraphics[scale=0.68,angle=90]{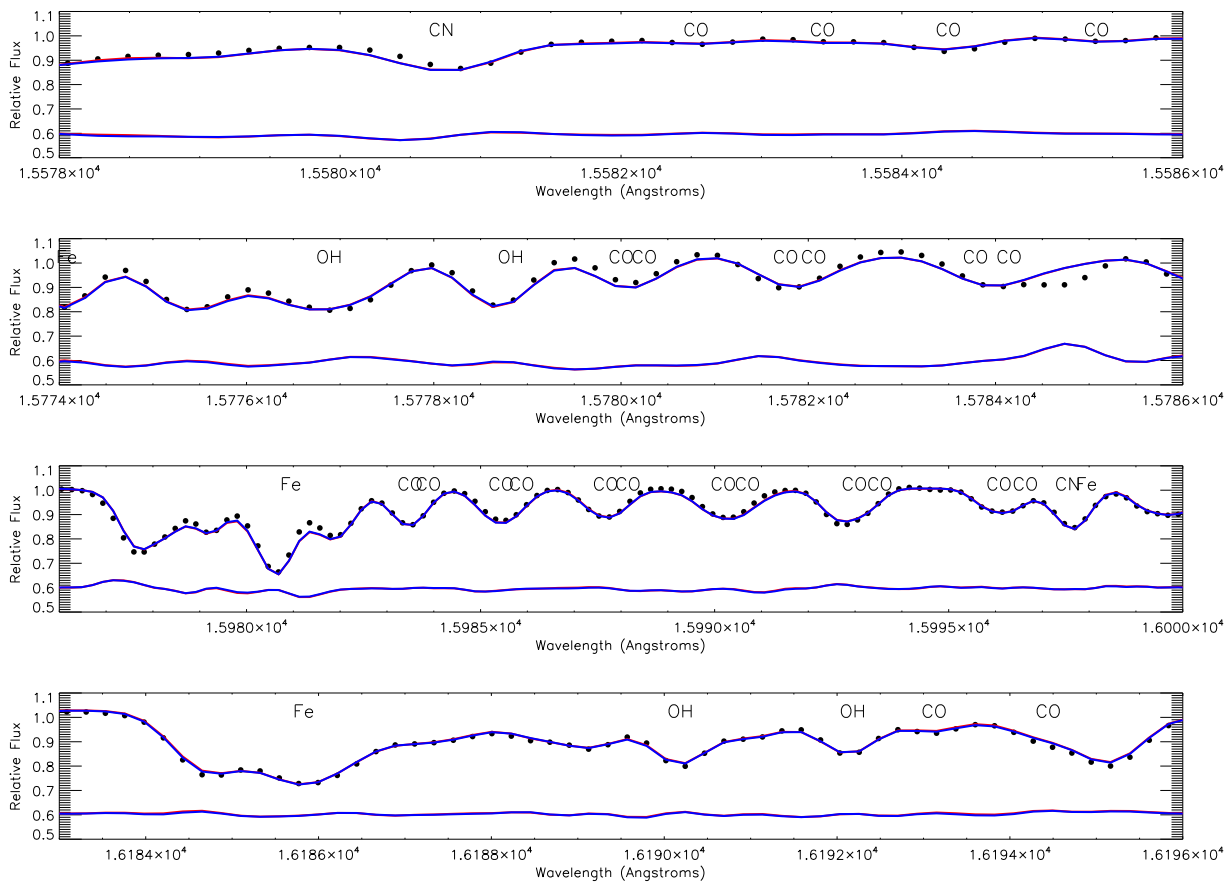}
\caption{Arcturus FTS observed spectrum (black dots) smoothed to the APOGEE
resolution ($R$ = 22,500), showing four $^{12}$C$^{16}$O
molecular windows/regions. The best fits obtained with the MARCS/Turbospectrum
(red line) and ATLAS9/ASS$\epsilon$T (blue line) synthetic spectra are also
shown. All spectra are expressed in air wavelengths. The residuals, computed as
flux(synthetic$-$observed)$+$0.60, are plotted at the bottom with the same color
code. The spectral features identified by Hinkle et al. (1995) are indicated
at the top.
\label{arcturus_C}}
\end{figure}

\begin{figure}
\epsscale{.80}
%\plotone{thesun_16500-16600.eps}
\begin{center}
\includegraphics[scale=0.5,angle=90]{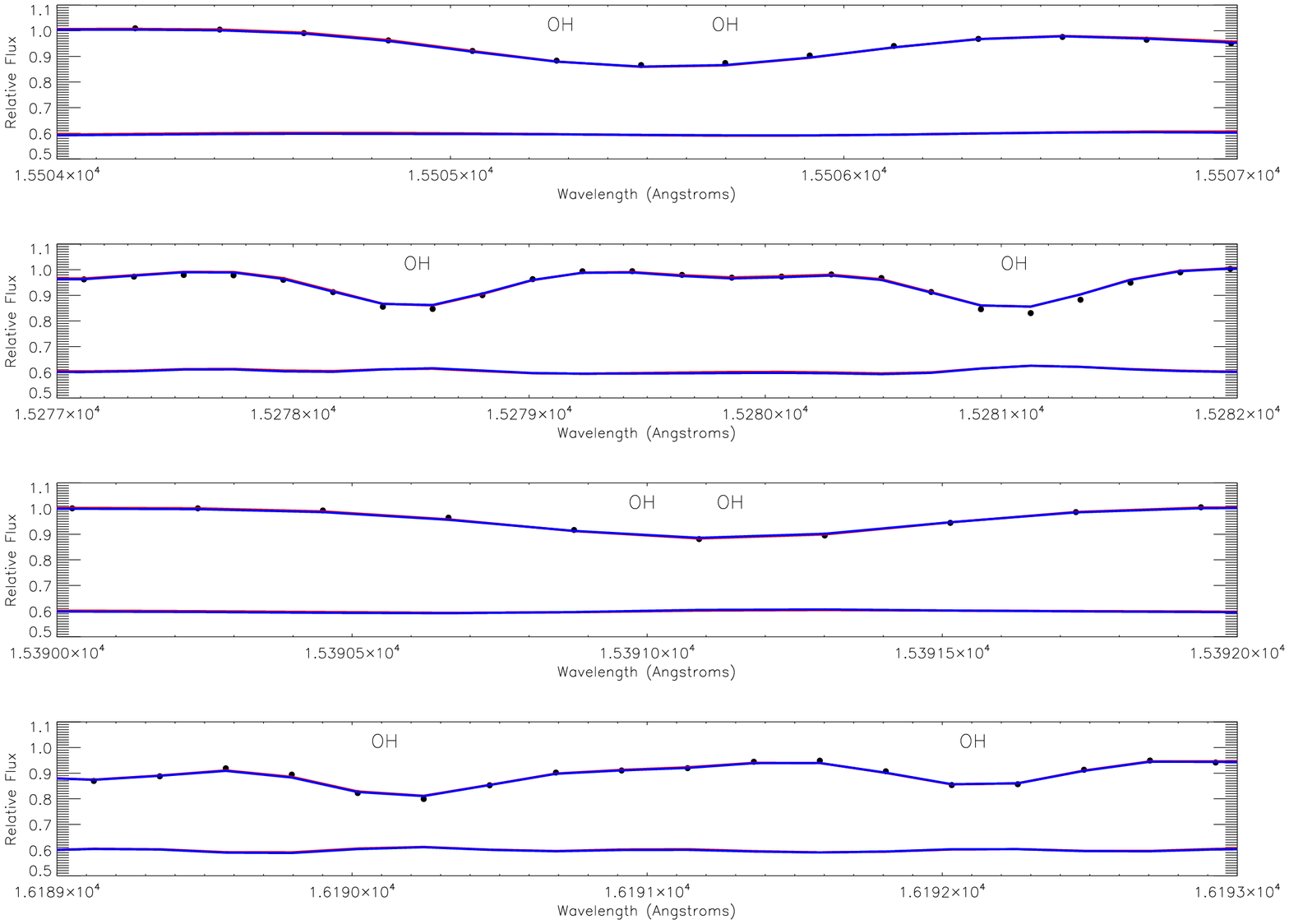}
\includegraphics[scale=0.5,angle=90]{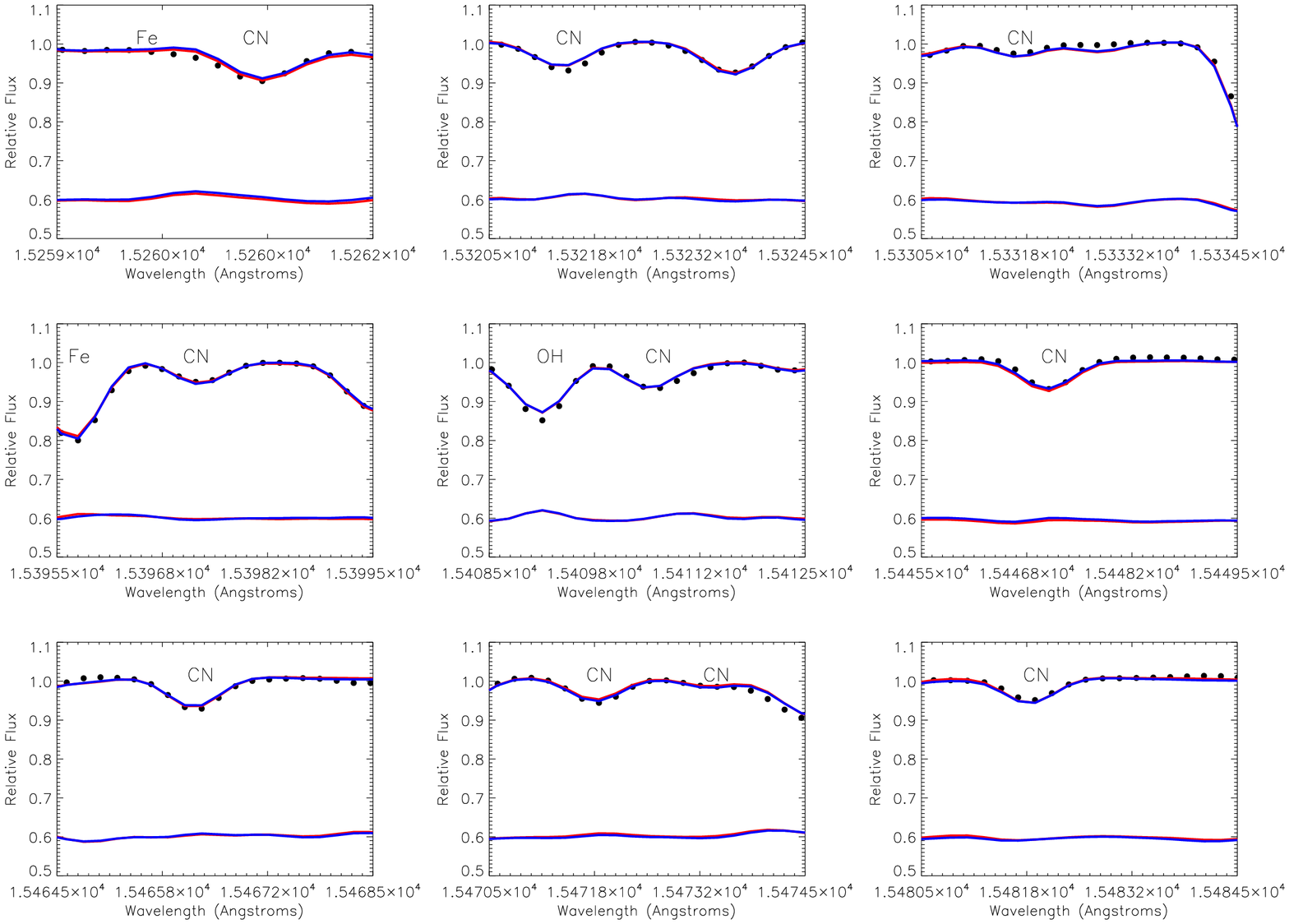}
\end{center}
\caption{Arcturus FTS observed spectrum (black dots) smoothed to the APOGEE
resolution ($R$ = 22,500) and the synthetic spectra in 
four $^{16}$OH molecular windows/regions (upper panel) and in nine
$^{12}$C$^{14}$N lines (lower panel). Symbols and colors as in Figure 8. 
The spectral features identified by Hinkle et al. (1995) are indicated
at the top.  
\label{arcturus_NO}}
\end{figure}

\begin{figure}
\epsscale{.80}
\plotone{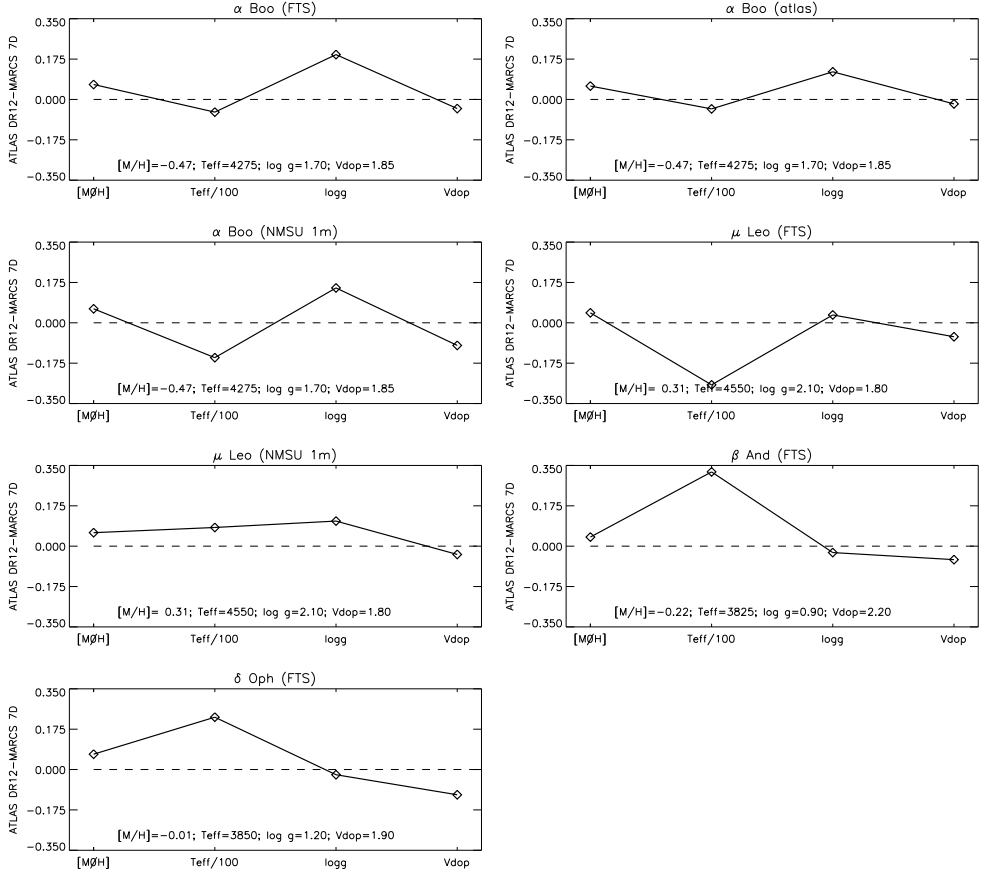}
\caption{Comparison between the atmospheric parameters derived by FERRE using
ATLAS9/ASS$\epsilon$T and MARCS/Turbospectrum for the calibration sample observed with the FTS. In order to have
all the values in the same scale, the $T\rm{_{eff}}$ value has been divided by
100. Moreover, the microturbulent velocity ($\xi$ $\equiv$ v$_{dop}$) is in
logarithmic scale, while the dashed line is the zero point. Note that stars
$\beta$ And and $\delta$ Oph have atmospheric parameters that correspond to a
hole in the MARCS/Turbospectrum grid (see text). \label{fig_fts_parameters}}
\end{figure}

\begin{figure}
\epsscale{.80}
\plotone{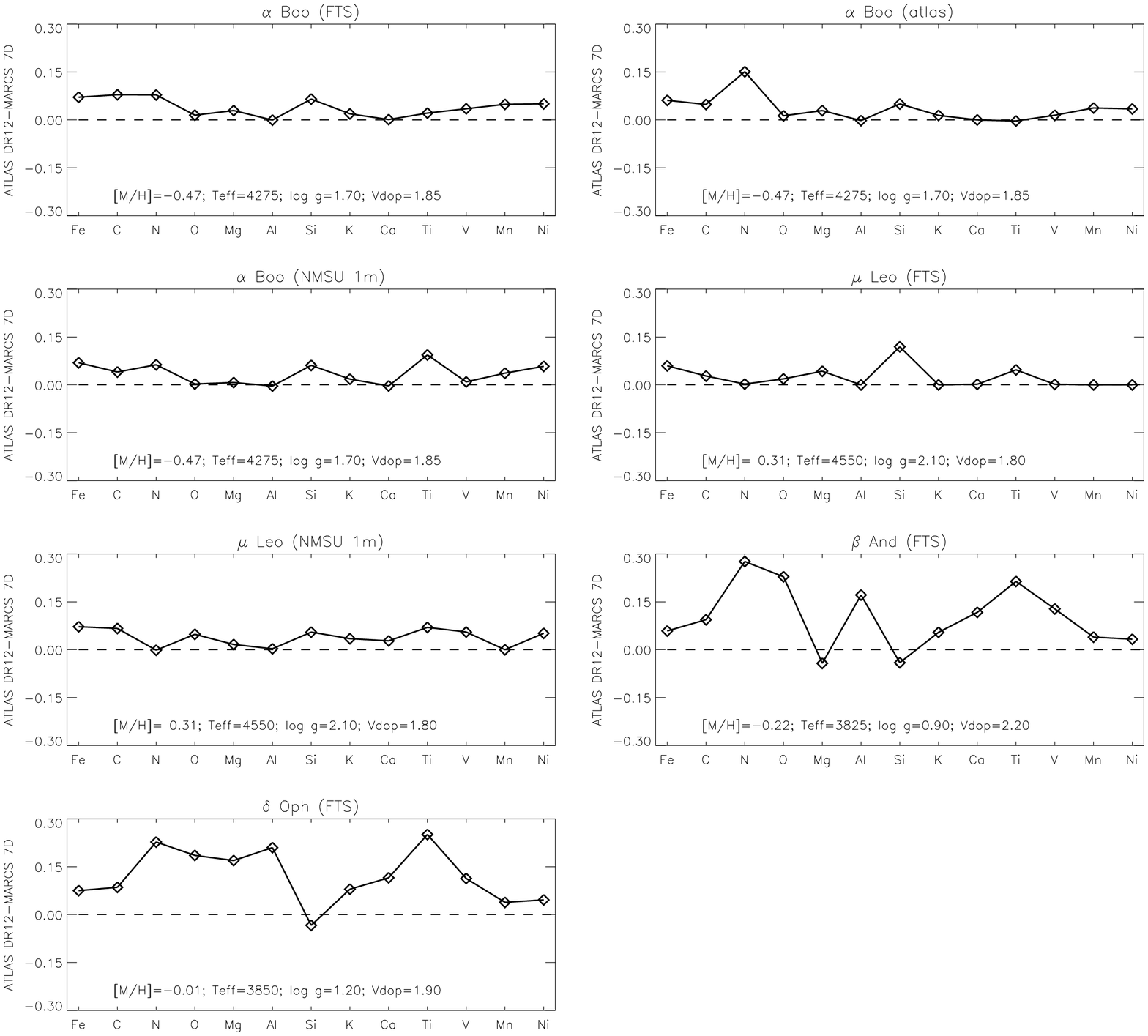}
\caption{Same as Fig.~\ref{fig_fts_parameters} but for the abundances of individual
elements. Note that stars $\beta$ And and $\delta$ Oph have atmospheric
parameters that correspond to a hole in the MARCS/Turbospectrum grid (see
text).\label{fig_fts_abundances}}
\end{figure}

\end{document}